\documentstyle[preprint,prd,aps,floats,epsf,tighten,amsmath]{revtex}

\begin{document}

\preprint{\vbox{
\hbox{JLAB-THY-00-14}
\hbox{hep-lat/0005002}
}}

\title{Domain Wall Fermions with Exact Chiral Symmetry}
\author{Robert G. Edwards}
\address{
Jefferson Lab,
12000 Jefferson Avenue,
MS 12H2,
Newport News, VA 23606, USA}
\author{Urs M. Heller}
\address{
CSIT, Florida State University, 
Tallahassee, FL 32306-4120, USA}

\date{May 7, 2000}

\maketitle

\begin{abstract} 
We show how the standard domain wall action can be simply modified to
allow arbitrarily exact chiral symmetry at finite fifth dimensional
extent. We note that the method can be used for both quenched and
dynamical calculations.  We test the method using smooth and
thermalized gauge field configurations.  We also make comparisons of
the performance (cost) of the domain wall operator for spectroscopy
compared to other methods such as the overlap-Dirac operator and find
both methods are comparable in cost.
\end{abstract}

\section{Introduction}

Recently a great deal of theoretical progress has been made in the construction
of lattice regularizations of fermions with good chiral properties
\cite{Kaplan,overlap,chiral_review}.
For use in practical numerical simulations, though, approximations to these
formulations are necessary. In the formulation using domain wall
fermions (DWF)~\cite{Kaplan,Shamir}, the extent of an auxiliary fifth dimension
has to be kept finite in numerical simulations while chiral symmetry
holds strictly only in the limit of infinite fifth dimension.
The violations of chiral symmetry are expected to be suppressed exponentially
in the extent of the fifth dimension~\cite{Shamir,finite_Ls}, but in practice
the coefficient in the exponent can be quite small~\cite{Columbia,dwf_problems}
and the suppression correspondingly slow.

In the case of overlap fermions there is no such problem in principle.
However, there is a problem of practicality: how to deal efficiently
with $\epsilon(H) = H / \sqrt{H^2}$, where $H$ is some auxiliary Hermitian
lattice Dirac operator for large negative mass, but free of doublers.
Most commonly, the Hermitian Wilson-Dirac operator $H_w$ is used.
So far, the best methods use rational polynomial approximations of
$\epsilon(H)$ rewritten as a sum over poles
\begin{eqnarray}
\epsilon(H) = \frac{H P(H^2)}{Q(H^2)} = H \left( c_0 + \sum_{k=1}^n
\frac{c_k}{H^2 + b_k} \right) ~.
\end{eqnarray}
Variants of this are Neuberger's polar decomposition~\cite{polar}
and the optimal rational polynomial approximation of
Ref.~\cite{EHN_practical,ParalComp}.
In the application of $\epsilon(H)$ on a vector $\psi$ the shifted
inversions $(H^2 + b_k)^{-1} \psi$ are done simultaneously with a
multi-shift CG inverter~\cite{Jegerlehner}. This is referred to as
the ``inner CG'', since typically another, ``outer'', CG method is
used to compute propagators or eigenvalues of the overlap Dirac operator
containing $\epsilon(H)$.

In these methods, the main problem is the difficulty of getting an
accurate approximation to $\epsilon(x)$ for small $x$\footnote{Using
the scaling invariance $\epsilon(sH) = \epsilon(H)$ for any positive
scale factor $s$ the upper range of the spectrum of $H$ can always be
scaled to be in the range of good accuracy of the approximation
to $\epsilon(x)$.}. As emphasized in Ref.~\cite{EHN_cond,ParalComp} accuracy
for the lower range of the spectrum of $H$ can always be enforced
by calculating a sufficient number of eigenvectors of $H$ with small
eigenvalues and computing $\epsilon(H)$ in the space spanned by these
eigenvectors exactly. The approximation is then only used after projecting
out these eigenvectors. This is referred to as ``projection''.

Small eigenvalues of $H_w$ occur quite frequently on (quenched) lattices
used in current simulations. Indeed, a non-zero density of eigenvalues
near zero has been found~\cite{EHN_rho0}. Projection is thus essential
in efficient implementations of the overlap Dirac operator. It is known
that the zero eigenvalues of $H_w$ correspond to unit eigenvalues
of the transfer matrix $T$ that describes propagation along the fifth
direction for domain wall fermions~\cite{overlap}. The left and right
handed physical fermions in the domain wall formulation occur along the
two boundaries in the fifth direction. A unit eigenvalue of the transfer
matrix then allows for unsuppressed interaction between left and right handed
fermions and thus to a breaking of chiral symmetry. Therefore the same
modes that make the approximation of $\epsilon(H)$ difficult to achieve
in overlap fermions, and need to be projected, are responsible for
the chiral symmetry breaking for domain wall fermions at finite $L_s$.

If we could project out the modes with near unit eigenvalue of the
transfer matrix and take their contribution with effectively infinite
$L_s$ we could obtain a domain wall fermion action with no chiral
symmetry breaking even at finite $L_s$. This would make domain wall fermions
with finite $L_s$ equivalent to overlap fermions. It would then become a
matter of computational efficiency as to which form is to be preferred
for numerical simulations. Based on ideas of Bori\c{c}i~\cite{Borici}
we show in section~\ref{sec:PDWF} how projection for domain wall fermions
can be achieved. In section~\ref{sec:results} we illustrate how projection
works, first on smooth instanton configurations and then on a configuration
from a quenched simulation. We discuss the degree of chiral symmetry
conservation that can be achieved and compare the performance (cost) of domain
wall fermion implementations with and without projection to overlap
fermions. We conclude the paper with a brief summary and some discussions in
section~\ref{sec:discussion}.

\section{Projected Domain Wall Fermion Actions}
\label{sec:PDWF}

Kaplan's method~\cite{Kaplan} realizes a single massless fermion
field through an infinite tower (or infinite number of flavors) of massive
fermion fields with a particular flavor structure. The flavor index can be
interpreted as an extra (here fifth) dimension, and the flavor structure as
a defect along the fifth dimension to which the massless fermions are
bound. Lattice calculations necessarily require a finite fifth
dimensional extent and lattice spacing, as well as regularizations for
the derivatives in the four dimensional space and in the fifth direction.
These regularizations are not unique. In particular, higher order terms
in the fifth dimensional lattice spacing can be added to the action,
as long as they do not change the ``defect structure''. Such changes
can affect the effective four dimensional theory by altering the
discretization effects in the four dimensional lattice spacing, but
not the continuum limit.

The finite fifth dimensional extent induces some chiral symmetry
violation. This can be eliminated by ``projecting'' an appropriate
number of violation inducing states.
We show how the projection method can be implemented for two different
five dimensional domain wall actions. The Wilson fermion action is
used for the four dimensional part of the action, but it should be
understood that other actions realizing a single massive fermion field
could be considered, and much of the following derivations do not
depend on the specific form chosen.

\subsection{Form of the actions}

To fix our notation/conventions, we write the usual, 4-d, Wilson-Dirac
operator as
\begin{eqnarray}
D_w(M) &=& (4+M) \delta_{x,y} - \frac{1}{2} \sum_{\mu=1}^4 \Bigl[
 (1 - \gamma_\mu) U_\mu(x) \delta_{x+\mu,y} +
 (1 + \gamma_\mu) U_\mu^\dagger(y) \delta_{x,y+\mu} \Bigr] \nonumber \\
 &=& \begin{pmatrix} B + M & C \\ -C^\dagger & B + M \end{pmatrix} .
\label{eq:D_w}
\end{eqnarray}
Here, with $\sigma_\mu = (\sigma_k, i {\bf 1})$,
\begin{eqnarray}
C &=& \frac{1}{2} \sum_{\mu=1}^4 \sigma_\mu \Bigl[ U_\mu(x)
 \delta_{x+\mu,y} - U_\mu^\dagger(y) \delta_{x,y+\mu} \Bigr] , \nonumber \\
B &=& \frac{1}{2} \sum_{\mu=1}^4 \Bigl[ 2 \delta_{x,y} -  U_\mu(x)
 \delta_{x+\mu,y} - U_\mu^\dagger(y) \delta_{x,y+\mu} \Bigr] .
\label{eq:C_B}
\end{eqnarray}
We are using a chiral basis: $\gamma_\mu = \begin{pmatrix} 0 & \sigma_\mu \\
\sigma_\mu^\dagger & 0 \end{pmatrix}$, $\gamma_5 = \begin{pmatrix} 1 & 0 \\
0 & -1 \end{pmatrix}$. We will also use the hermitian Wilson-Dirac operator
\begin{eqnarray}
H_w(M) = \gamma_5 D_w(M) = \begin{pmatrix} B + M & C \\
        C^\dagger & -B - M \end{pmatrix} .
\label{eq:H_w}
\end{eqnarray}
Usually, we will use a large negative mass here, but often omit the
argument from $D_w$ and $H_w$.

In this notation, the usual domain wall fermion action of Shamir~\cite{Shamir}
reads, with an explicit fifth dimensional lattice spacing $a_5$
for the hopping term --- the 4-d lattice spacing $a$ is kept fixed at $a=1$
throughout ---,
\begin{eqnarray}
S_{DW} = - \bar \Psi D^{(5)}_{DW} \Psi = - \sum_{i=1}^{L_s} \bar \Psi_i
 \Bigl\{ \bigl[a_5 D_w(-M) + 1 \bigr] \Psi_i - P_- \Psi_{i+1} -
 P_+ \Psi_{i-1} \Bigr\}
\label{eq:S_DW}
\end{eqnarray}
where the extent of the fifth dimension, $L_s$ has to be taken as even.
The gauge fields in $D_w(-M)$ are independent of the fifth coordinate,
and the fermion fields satisfy the boundary conditions in the fifth
direction:
\begin{eqnarray}
P_- \Psi_{L_s+1} = -m P_- \Psi_1 ,  \qquad P_+ \Psi_0 = -m P_+ \Psi_{L_s} .
\label{eq:bc}
\end{eqnarray}
$P_\pm$ are the chiral projectors, $P_\pm = \frac{1}{2} ( 1 \pm \gamma_5)$.
$0 \le m < 1$ is proportional to the quark mass (for small $m$).
The 4-d physical fermion degrees of freedom are identified with the fields
at the boundaries as
\begin{eqnarray}
q^R = P_+ \Psi_{L_s} = \Psi_{L_s}^R ,&& \quad  q^L = P_- \psi_1 = \Psi_1^L ,
 \nonumber \\
\bar q^R = \bar \Psi_{L_s} P_- = \bar \Psi_{L_s}^R ,&& \quad \bar q^L =
\bar \psi_1 P_+ = \bar \Psi_1^L .
\label{eq:light}
\end{eqnarray}

As will be shown below, projection of low-lying eigenvalues of $-\log T$,
with $T$ the transfer matrix along the fifth direction, can be achieved
by introducing an additional term in (\ref{eq:S_DW}) so that the
complete domain wall fermion action reads
\begin{eqnarray}
S_{DWP} = - \bar \Psi D^{(5)}_{DWP} \Psi = &-& \sum_{i=1}^{L_s}
 \bar \Psi_i \Bigl\{ \bigl[a_5 D_w(-M) + 1 \bigr] \Psi_i -
 P_- \Psi_{i+1} - P_+ \Psi_{i-1} \Bigr\} \nonumber \\
 &+& \qquad \bar \Psi_1 \hat A(m) \bigr[ P_- \Psi_1 + P_+ \Psi_{L_s} \bigl] .
\label{eq:S_DWP}
\end{eqnarray}

Bori\c{c}i~\cite{Borici} introduced an interesting variation of the
domain wall fermion action, which differs from the usual one by terms
of order ${\cal O}(a_5)$. He called it (5-d) truncated overlap action,
since its effective 4-d action for the light fermions is just the
polar decomposition approximation of order $L_s/2$ to the overlap Dirac
operator introduced by Neuberger~\cite{polar}. In our notation, and
introducing again an additional term to be used for projecting
low-lying eigenvalues, Bori\c{c}i's variant reads
\begin{eqnarray}
S_{DW'} = - \bar \Psi D^{(5)}_{DW'} \Psi = - \sum_{i=1}^{L_s} \bar \Psi_i
 && \Bigl\{ \bigr[a_5 D_w(-M) + 1 \bigl] \Psi_i +
 \bigr[a_5 D_w(-M) - 1 \bigl] P_- \Psi_{i+1} + \nonumber \\
 && ~~ \bigr[a_5 D_w(-M) - 1 \bigl] P_+ \Psi_{i-1} \Bigr\}
 + \bar \Psi_1 \hat A(m) \bigr[ P_- \Psi_1 + P_+ \Psi_{L_s} \bigl] .
\label{eq:S_B}
\end{eqnarray}
The boundary conditions in the fifth direction remain as in (\ref{eq:bc}).
The kernel of the 5-d operator is, including the boundary conditions,
\begin{eqnarray}
D^{(5)}_{DW'} = \begin{pmatrix}
D_+ - \hat{A}P_- & D_-P_- & 0 & 0&\cdots & 0 & 0 &
-m D_-P_+ - \hat{A} P_+ \cr
D_-P_+ & D_+ & D_-P_- & 0&\cdots & 0 & 0 & 0 \cr
0 & D_-P_+ & D_+ & D_-P_- &\cdots & 0 & 0 & 0 \cr
\vdots & \vdots & \vdots & \vdots & \vdots & \vdots & \vdots & \vdots \cr
0 & 0 & 0 & 0 & \cdots & D_-P_+ & D_+ & D_-P_- \cr
-m D_-P_- & 0 & 0 & 0 & \cdots & 0 & D_-P_+ & D_+ \cr
\end{pmatrix}
\label{eq:D_5d_matrix}
\end{eqnarray}
where
\begin{eqnarray}
D_\pm = a_5 D_w(-M) \pm 1 ~.
\end{eqnarray}
The kernel of the standard domain wall operator, but including the term
to be used for projection, $D^{(5)}_{DWP}$, is simply obtained by the
replacement $D_- \to -1$.

We now integrate out the heavy fermion degrees of freedom to arrive at
a 4-d Dirac operator describing the light fermions.
Following Ref.~\cite{Borici} we define ${\cal P}$ by
\begin{eqnarray}
{\cal P}_{jk} = \begin{cases}
 P_- \delta_{j,k} + P_+ \delta_{j+1,k} & \text{for $j = 1, \dots, L_s-1$} \\
 P_- \delta_{L_s,k} + P_+ \delta_{1,k} & \text{for $j = L_s$} ~. \end{cases}
\end{eqnarray}
This has an inverse ${\cal P}^{-1} = {\cal P}^\dagger$, given by
\begin{eqnarray}
{\cal P}^{-1}_{jk} = \begin{cases}
 P_- \delta_{j,k} + P_+ \delta_{j-1,k} & \text{for $j = 2, \dots, L_s$} \\
 P_- \delta_{1,k} + P_+ \delta_{L_s,k} & \text{for $j = 1$} ~. \end{cases}
\end{eqnarray}

Next, we introduce $\chi_i$'s through $\Psi_i = ({\cal P} \chi)_i$
and define 4-d operators $Q_\pm$ as
\begin{eqnarray}
Q_\pm = \begin{cases}
 a_5 H_w P_\pm \pm 1 & \text{for the standard domain wall action} \\
 a_5 H_w \pm 1 & \text{for Bori\c{c}i's domain wall action} ~. \end{cases}
\label{eq:Q}
\end{eqnarray}
Then, using $\gamma_5 P_+ = P_+$ and $\gamma_5 P_- = - P_-$ as well as
the boundary conditions on the fermion fields, we can rewrite both domain
wall actions as
\begin{eqnarray}
\label{eq:S_5d2}
S^{(5)} = &-& \Bigl\{ \bar \Psi_1 \gamma_5 \bigl[ Q_-
 (P_- -m P_+) \chi_1 - \gamma_5 \hat A \chi_1 + Q_+ \chi_2 \bigr] \\
 &+& \quad \sum_{i=2}^{L_s-1} \bar \Psi_i \gamma_5
 \bigl[ Q_- \chi_i + Q_+ \chi_{i+1} \bigr] +
 \bar \Psi_{L_s} \gamma_5 \bigl[ Q_- \chi_{L_s} + Q_+ (P_+ -m P_-) \chi_1 \bigr]
 \Bigr\} . \nonumber
\end{eqnarray}

Next, we introduce $\bar \Psi_i = \bar \chi_i Q_-^{-1} \gamma_5$ and
\begin{eqnarray}
T^{-1} = - Q_-^{-1} Q_+ ~.
\label{eq:T}
\end{eqnarray}
This change of variables would give rise to a Jacobian in a dynamical
simulation. However, it would be exactly canceled by the Jacobian of
this transformation for the pseudo-fermion fields which are needed to
cancel the bulk contribution from the 5-d fermions. The action for the
pseudo-fermions is obtained by the replacement $m \to 1$ from the
fermion action. The Jacobians cancel since integration of fermions
(Grassman fields) acts like differentiation. We note that for the
standard domain wall action $Q_\pm$ do not commute and the ordering in
(\ref{eq:T}) is important. From (\ref{eq:H_w}) we find
\begin{eqnarray}
T^{-1} = \begin{pmatrix} 1 & a_5 C \frac{1}{\tilde B} \cr
 0 & \frac{1}{\tilde B} \cr \end{pmatrix} \begin{pmatrix} {\tilde B} & 0 \cr
 a_5 C^\dagger & 1 \end{pmatrix} = \begin{pmatrix}
 \tilde B + a_5^2 C \frac{1}{\tilde B} C^\dagger & a_5 C \frac{1}{\tilde B} \cr
 a_5 \frac{1}{\tilde B} C^\dagger & \frac{1}{\tilde B} \cr \end{pmatrix} ~,
\label{eq:Tinv_DW}
\end{eqnarray}
where $\tilde B = 1 + a_5(B - M)$, and thus
\begin{eqnarray}
T = \begin{pmatrix} \frac{1}{\tilde B} & 0 \cr
 - a_5 C^\dagger \frac{1}{\tilde B} & 1 \cr \end{pmatrix} \begin{pmatrix}
 1 & - a_5 C \cr 0 & {\tilde B} \cr \end{pmatrix} = \begin{pmatrix}
 \frac{1}{\tilde B} & - a_5 \frac{1}{\tilde B} C \cr
 - a_5 C^\dagger \frac{1}{\tilde B} & a_5^2 C^\dagger \frac{1}{\tilde B} C
 + \tilde B \cr \end{pmatrix} ~.
\label{eq:T_DW}
\end{eqnarray}
$T$ is the usual domain wall fermion transfer matrix~\cite{overlap}
in our conventions.

For later use it will be convenient to introduce a 4-d Hamiltonian
$H_T$~\cite{Borici} such that
\begin{eqnarray}
T^{-1} = \frac{1 + a_5 H_T}{1 - a_5 H_T} ~.
\label{eq:T_HT}
\end{eqnarray}
For Bori\c{c}i's domain wall action we have simply $H_T = H_w$, while
for the standard domain wall action one finds
\begin{eqnarray}
H_T = \frac{1}{2 + a_5 H_w \gamma_5} H_w =
 H_w \frac{1}{2 + a_5 \gamma_5 H_w} ~.
\label{eq:H_T}
\end{eqnarray}

In terms of the new fields $\bar \chi$ and $\chi$ the 5-d actions become
\begin{eqnarray}
S^{(5)} = - \bar \chi D^{(5)}_\chi \chi =
 &-& \biggl\{ \bar \chi_1 \bigl[ (P_- -m P_+) \chi_1 -
 Q_-^{-1} \gamma_5 \hat A \chi_1 - T^{-1} \chi_2 \bigr] \nonumber \\
 &+& \quad \sum_{i=2}^{L_s-1} \bar \chi_i \bigl[ \chi_i-
 T^{-1} \chi_{i+1} \bigr]
 + \bar \chi_{L_s} \bigl[ \chi_{L_s} - T^{-1} (P_+ -m P_-) \chi_1 \bigr] \biggr\} .
\label{eq:S_5d3}
\end{eqnarray}

Now, we integrate out, in succession, $\chi_{L_s}, \bar \chi_{L_s}$, $\chi_{L_s-1},
\bar \chi_{L_s-1}$, $\dots$, $\chi_2, \bar \chi_2$. For this we use, at
the first step,
\begin{eqnarray}
 && \bar \chi_{L_s} \chi_{L_s} - \bar \chi_{L_s-1} T^{-1} \chi_{L_s} -
 \bar \chi_{L_s} T^{-1} (P_+ -m P_-) \chi_1 = \nonumber \\
 && \bigl[ \bar \chi_{L_s} - \bar \chi_{L_s-1} T^{-1} \bigr]
 \bigl[ \chi_{L_s} - T^{-1} (P_+ -m P_-) \chi_1 \bigr] -
 \bar \chi_{L_s-1} T^{-2} (P_+ -m P_-) \chi_1
\nonumber
\end{eqnarray}
and at the $(L_s-i)$-th step
\begin{eqnarray}
\label{eq:int_i}
 && \bar \chi_{i+1} \chi_{i+1} - \bar \chi_i T^{-1}
 \chi_{i+1} - \bar \chi_{i+1} T^{-L_s+i} (P_+ -m P_-) \chi_1 = \\
 && \bigl[ \bar \chi_{i+1} - \bar \chi_i T^{-1} \bigr]
 \bigl[ \chi_{i+1} - T^{-L_s+i} (P_+ -m P_-) \chi_1 \bigr] -
 \bar \chi_i T^{-L_s+i-1} (P_+ -m P_-) \chi_1 . \nonumber
\end{eqnarray}
With a change of variables $\chi_{i+1}^\prime = \chi_{i+1} - T^{-L_s+i}
(P_+ -m P_-) \chi_1$ and $\bar \chi_{i+1}^\prime = \bar \chi_{i+1} -
\bar \chi_i T^{-1}$ the integration over $\chi_{i+1}^\prime,
\bar \chi_{i+1}^\prime$ is trivial, giving a factor 1.
At the end, we arrive at the 4-d action for $\chi_1, \bar \chi_1$
\begin{eqnarray}
S^{(4)} = - \bar \chi_1 \bigl[ (P_- -m P_+) - T^{-L_s} (P_+ -m P_-)
 - Q_-^{-1} \gamma_5 \hat A \bigr] \chi_1
 = - \bar \chi_1 D^{(4)}(m) \chi_1 .
\label{eq:S_4d}
\end{eqnarray}
The kernel is
\begin{eqnarray}
D^{(4)}(m) &=& (P_- -m P_+) - T^{-L_s} (P_+ -m P_-) - Q_-^{-1}
 \gamma_5 \hat A \nonumber \\
\label{eq:D_4d}
 &=& -\left[ \frac{1+m}{2} \left(T^{-L_s} + 1 \right) \gamma_5 +
 \frac{1-m}{2} \left(T^{-L_s} -1 \right) + Q_-^{-1} \gamma_5 \hat A \right] \\
 &=& - \Bigl[ \left(T^{-L_s} + 1 \right) \gamma_5 \Bigr] \times
 \left[ \frac{1+m}{2} + \frac{1-m}{2} \gamma_5 \frac{T^{-L_s} -1}{T^{-L_s} + 1}
 + \gamma_5 \left[ Q_-(T^{-L_s} + 1) \right]^{-1} \gamma_5 \hat A \right] ~.
\nonumber
\end{eqnarray}

Finally, integrating out $\chi_1, \bar \chi_1$ and dividing by the
pseudo-fermion determinant, obtained by the substitution $m \to 1$
and requiring $\hat A(m=1) = 0$, we find
\begin{eqnarray}
\frac{\det D^{(5)}(m)}{\det D^{(5)}(1)}
 = \frac{\det D^{(4)}(m)}{\det D^{(4)}(1)}
 = \det \left\{ \left[ D^{(4)}(1) \right]^{-1} D^{(4)}(m) \right\} .
\label{eq:5d_4d_det}
\end{eqnarray}
Now, from eq.~(\ref{eq:D_4d}), $D^{(4)}(1) = - \left(T^{-L_s} + 1 \right)
\gamma_5$ and thus we get
\begin{eqnarray}
\left[ D^{(4)}(1) \right]^{-1} D^{(4)}(m) =
 \frac{1}{2} \Biggl[1+m + (1-m) \gamma_5 \frac{T^{-L_s} -1}{T^{-L_s} + 1}
 + 2 \gamma_5 \left[ Q_-(T^{-L_s} + 1) \right]^{-1} \gamma_5 \hat A(m) \Biggr] ~.
\label{eq:D_T}
\end{eqnarray}
Using (\ref{eq:T_HT}) we note that
\begin{eqnarray}
\frac{T^{-L_s} -1}{T^{-L_s} + 1} = \frac{(1+a_5 H_T)^{L_s} - (1-a_5 H_T)^{L_s}}
 {(1+a_5 H_T)^{L_s} + (1-a_5 H_T)^{L_s}} = \varepsilon_{L_s/2}(a_5 H_T)
\end{eqnarray}
where $\varepsilon_n(x)$ is Neuberger's polar decomposition approximation
to $\epsilon(x)$. Ignoring the term with $\hat A(m)$ for the moment, we see
that
\begin{eqnarray}
\left[ D^{(4)}(1) \right]^{-1} D^{(4)}(m) =
\frac{1}{2} \Bigl[1+m + (1-m) \gamma_5 \varepsilon_{L_s/2}(a_5 H_T) \Bigr]
 = D_{tov}(m) ~.
\label{eq:D_TOV}
\end{eqnarray}
This is just Neuberger's polar decomposition approximation \cite{polar}
to the overlap Dirac operator for auxiliary Hamiltonian $H_T$, which we
denote by $D_{tov}$, ``truncated overlap.'' For $L_s \to \infty$ it becomes
the exact overlap Dirac operator $D_{ov}$~\cite{Herbert}.
Alternatively, we can write
\begin{eqnarray}
\frac{T^{-L_s} -1}{T^{-L_s} + 1} = \tanh\left(-\frac{L_s}{2} \log |T| \right)
\label{eq:dwf_logT}
\end{eqnarray}
and eq.~(\ref{eq:D_T}), still ignoring the term with $\hat A(m)$, becomes the
effective 4-d Dirac operator that Neuberger derived for domain wall fermions
in \cite{Herbert_DWF}.
Here we have used that $L_s$ is even and written the formula in terms
of the absolute value $|T|$ to indicate that everything remains well
defined when an eigenvalue of $T$ becomes negative.

We now can use $\hat A(m)$ to project out low-lying eigenvectors, $v_i$, of
the auxiliary Hamiltonian $H_T$
for which $\varepsilon_{L_s/2}$ with finite $L_s$ is not a sufficiently accurate
approximation to $\epsilon(x)$. Let
\begin{eqnarray}
H_T v_i = \lambda_i v_i, \qquad T v_i = T_i v_i, \qquad
\hat{P}_i = v_i v_i^\dagger ~,
\end{eqnarray}
where, from eq.~(\ref{eq:T_HT}), 
\begin{eqnarray}
T_i = \frac{1 - a_5 \lambda_i}{1 + a_5 \lambda_i} ~.
\label{eq:T_i}
\end{eqnarray}
The projection can then be achieved by setting
\begin{eqnarray}
\hat A(m) = (1-m) \gamma_5 Q_- \sum_i g_i \hat{P}_i ,
\label{eq:hat_A_dwf}
\end{eqnarray}
with
\begin{eqnarray}
g_i &=& \frac{1}{2} \left[-\left(T_i^{-L_s} -1 \right) +
 \left(T_i^{-L_s} + 1\right)\epsilon(a_5 \lambda_i)\right] ~.
\label{eq:g_i}
\end{eqnarray}
Note that $\hat A(m)$ vanishes for $m=1$, as required for the
pseudo-fermions.
With this $\hat A(m)$ we find, instead of (\ref{eq:D_TOV}),
\begin{eqnarray}
&& \left[ D^{(4)}(1) \right]^{-1} D^{(4)}(m) = D_{ov}(m) = \nonumber \\
&& \qquad \frac{1}{2} \biggl\{1+m + (1-m) \gamma_5 \Bigl[
 \varepsilon_{L_s/2}(a_5 H_T) \Bigl(1 - \sum_i \hat{P}_i \Bigr) +
 \sum_i \epsilon(a_5 \lambda_i) \hat{P}_i \Bigr] \biggr\} ~.
\label{eq:D_ov_proj}
\end{eqnarray}

For Bori\c{c}i's variant of the domain wall action, where $H_T=H_w$
commutes with $Q_\pm$ the ``projection operator'' $\hat A(m)$ can be
simplified to
\begin{eqnarray}
\hat A(m) = (1-m) \gamma_5 \sum_i f_i \hat{P}_i ,
\end{eqnarray}
with
\begin{eqnarray}
f_i &=& \frac{1}{2}(a_5\lambda_i -1) \left[-\left(T_i^{-L_s} -1 \right) +
 \left(T_i^{-L_s} + 1\right)\epsilon(a_5 \lambda_i)\right]\nonumber\\
  &=& \begin{cases}
       (a_5\lambda_i -1)&\text{for $\lambda_i > 0$}\\
       \frac{(1 + a_5\lambda_i)^{L_s}}{(1- a_5\lambda_i)^{L_s-1}}&
       \text{for $\lambda_i < 0$}~.
      \end{cases}
\label{eq:f_i}
\end{eqnarray}

\subsection{Propagator}

Our next step is to relate the 4-d overlap propagator to the
5-d propagator of the corresponding domain wall fermion action. In all
steps below $\hat A(m)$ for the eigenvalue projections is included. From
(\ref{eq:D_ov_proj}) we find, obviously,
\begin{eqnarray}
D_{ov}^{-1}(m) = \left[ D^{(4)}(m) \right]^{-1} D^{(4)}(1) .
\label{eq:D_ov_inv}
\end{eqnarray}
To connect this to the 5-d theory, we consider, motivated by the fact
that the light 4-d fermion is $q = ({\cal P}^{-1} \Psi)_1 = \chi_1$,
\begin{eqnarray}
X &=& \left\{ {\cal P}^{-1} \left[ D^{(5)}(m) \right]^{-1}
 D^{(5)}(1) {\cal P} \right\}_{11} \nonumber \\
\label{eq:X}
 &=& \frac{1}{Z} \int \prod_i d\Psi_i d\bar \Psi_i \sum_k
 ({\cal P}^{-1} \Psi)_1 \bar \Psi_k \left[D^{(5)}(1) {\cal P} \right]_{k1}
 {\rm e}^{-S^{(5)}} \\
 &=& \frac{1}{Z^\prime} \int \prod_i d\chi_i d\bar \chi_i \sum_k
 \chi_1 \bar \chi_k Q_-^{-1} \gamma_5 \left[D^{(5)}(1)
 {\cal P} \right]_{k1} {\rm e}^{-S^{(5)}} \nonumber \\
 &=& \frac{1}{Z^\prime} \int \prod_i d\chi_i d\bar \chi_i \sum_k
 \chi_1 \bar \chi_k \left[ D^{(5)}_\chi(1) \right]_{k1} {\rm e}^{-S^{(5)}} ~.
 \nonumber
\end{eqnarray}
Here we have used the 5-d $D^{(5)}_\chi = Q_-^{-1} \gamma_5 D^{(5)} {\cal P}$
introduced in eq.~(\ref{eq:S_5d3}).
Now, we integrate out, in succession, $\chi_{L_s}, \bar \chi_{L_s}$,
$\chi_{L_s-1}, \bar \chi_{L_s-1}$, $\dots$, $\chi_2, \bar \chi_2$. From the
transformations (\ref{eq:int_i}) we see that $\bar \chi_k \rightarrow
\bar \chi_1 T^{-k+1}$ in the process. Therefore, we obtain
\begin{eqnarray}
X = \frac{1}{Z^\prime} \int d\chi_1 d\bar \chi_1 \chi_1 \bar \chi_1
 \sum_k T^{-k+1} \left[ D^{(5)}_\chi(1) \right]_{k1}
 {\rm e}^{\bar \chi_1 D^{(4)}(m) \chi_1} .
\end{eqnarray}
But from (\ref{eq:S_5d3}) we see that
\begin{eqnarray}
\sum_k T^{-k+1} \left[ D^{(5)}_\chi(1) \right]_{k1} = (P_- - P_+) -
 T^{-L_s} (P_+ - P_-) = D^{(4)}(1) ~,
\label{eq:pseudo_term}
\end{eqnarray}
where we used the fact that $\hat A(1) = 0$.
Thus we finally obtain
\begin{eqnarray}
X = \left\{ {\cal P}^{-1} \left[ D^{(5)}(m) \right]^{-1}
 D^{(5)}(1) {\cal P} \right\}_{11}
 = \left[ D^{(4)}(m) \right]^{-1} D^{(4)}(1) = D_{ov}^{-1}(m) ~.
\label{eq:D_ov5d_inv}
\end{eqnarray}

To solve
\begin{eqnarray}
D_{ov}(m) \psi = b
\label{eq:d_prop}
\end{eqnarray}
we introduce $\tilde b = (b,0,\dots,0)^T$ and solve~\cite{Borici}
\begin{eqnarray}
D^{(5)}(m) \phi = D^{(5)}(1) {\cal P} \tilde b .
\label{eq:5d_prop}
\end{eqnarray}
$\psi$ is then obtained as
\begin{eqnarray}
\psi = \left( {\cal P}^{-1} \phi \right)_1 .
\label{eq:prop_soln}
\end{eqnarray}

The physical propagator has a contact term subtracted~\cite{EHN_cond},
and we arrive at the general form for both auxiliary Hamiltonians
considered here, with and without projection, and also valid for
finite $L_s$,
\begin{eqnarray}
{\tilde D_{ov}}^{-1}(m) 
  = \frac{1}{1-m} \left[D_{ov}^{-1}(m) - 1 \right]
  = \frac{1}{1-m}\left[\left\{ {\cal P}^{-1} \left[ D^{(5)}(m) \right]^{-1}
 D^{(5)}(1) {\cal P} \right\}_{11} - 1\right]~.
\label{eq:4d5d_prop}
\end{eqnarray}

\subsection{Relation of the 5-d and 4-d operators}

So far, we have related $D_{ov}^{-1}(m)$ and $\det D_{ov}(m)$, or their
truncated versions, to 5-d operators, see (\ref{eq:5d_4d_det}),
(\ref{eq:D_TOV}) and (\ref{eq:D_ov_proj}) and finally (\ref{eq:D_ov5d_inv}).
{\it E.g.} for eigenvalue calculations it would be useful to establish
a similar connection for $D_{ov}(m)$ or $D_{tov}(m)$. Consider
$ D^{(5)}_\chi(m)$ introduced in (\ref{eq:S_5d3}). We find
\begin{eqnarray}
&& \left[ D^{(5)}_\chi(m) - D^{(5)}_\chi(1) \right]_{ij} = \\
 && \left[ (1-m) P_+ - Q_-^{-1} \gamma_5 \hat A(m) \right]
 \delta_{i1} \delta_{j1} - T^{-1} (1-m) P_- \delta_{iL_s} \delta_{j1} ~.
 \nonumber
\end{eqnarray}
Thus
\begin{eqnarray}
&& \left[ D^{(5)}_\chi(1) \right]^{-1} D^{(5)}_\chi(m) = \\
 && 1 + \left[ D^{(5)}_\chi(1) \right]^{-1}_{i1}
 \left[ (1-m) P_+ - Q_-^{-1} \gamma_5 \hat A(m) \right] \delta_{j1}
 - (1-m) \left[ D^{(5)}_\chi(1) \right]^{-1}_{iL_s} T^{-1} P_- \delta_{j1} ~.
 \nonumber
\end{eqnarray}
So only the first column (of 4-d blocks) is different from the 5-d unit
matrix. Let's call the entries in the first column $X_1, X_2, \dots , X_{L_s}$.
Similarly, its inverse is
\begin{eqnarray}
&& \left[ D^{(5)}_\chi(m) \right]^{-1} D^{(5)}_\chi(1) = \\
 && 1 - \left[ D^{(5)}_\chi(m) \right]^{-1}_{i1}
 \left[ (1-m) P_+ - Q_-^{-1} \gamma_5 \hat A(m) \right] \delta_{j1}
 + (1-m) \left[ D^{(5)}_\chi(m) \right]^{-1}_{iL_s} T^{-1} P_- \delta_{j1} ~.
 \nonumber
\end{eqnarray}
Again, only the first column is different from the 5-d unit matrix. Let's
call the entries in the first column $Y_1, Y_2, \dots , Y_{L_s}$.
Since these 5-d matrices are the inverse of each other, their product
is the 5-d unit matrix, {\it i.e.}
\begin{eqnarray}
Y_1 X_1 = 1  &\Rightarrow&  Y_1 = (X_1)^{-1} \nonumber \\
Y_2 X_1 + X_2 = 0  &\Rightarrow&  Y_2 = - X_2 (X_1)^{-1} \nonumber \\
\vdots   & & \vdots \\
Y_j X_1 + X_j = 0  &\Rightarrow&  Y_j = - X_j (X_1)^{-1} \quad {\rm for}~
j=1, \dots, L_s ~.
\end{eqnarray}
The first equation, $Y_1 = (X_1)^{-1}$ establishes the relation we are looking
for
\begin{eqnarray}
D_{ov}(m) = \left\{ {\cal P}^{-1} \left[ D^{(5)}(1) \right]^{-1}
 D^{(5)}(m) {\cal P} \right\}_{11} ~.
\label{eq:op5d4d}
\end{eqnarray}

Note that the components of the 5-d propagator used for the inverse
of the overlap Dirac operator in eq.~(\ref{eq:D_ov5d_inv}) do not
correspond to the physical quark propagator $\langle q \bar q \rangle$ as
obtained from domain wall fermions. The reason is that the overlap
propagator still needs a subtraction of a contact term and a multiplicative
normalization~\cite{EHN_cond}. We consider here the standard domain wall
action without projection, and denote the corresponding 4-d Dirac
operator $D_{tov}(m;H_T)$. To make the connection between
the subtracted 4-d propagator and $\langle q \bar q \rangle$
explicit, we write the 1 for the subtraction as
\begin{eqnarray}
1 = \left\{ {\cal P}^{-1} \left[ D^{(5)}_{DW}(m) \right]^{-1}
 D^{(5)}_{DW}(m) {\cal P} \right\}_{11}
\end{eqnarray}
and thus
\begin{eqnarray}
D_{tov}^{-1}(m;H_T) - 1 = \left\{ {\cal P}^{-1} \left[ D^{(5)}_{DW}(m) \right]^{-1}
 \left[ D^{(5)}_{DW}(1) - D^{(5)}_{DW}(m) \right] {\cal P} \right\}_{11} .
\end{eqnarray}
{}From the matrix for $D^{(5)}_{DW}$, eq.~(\ref{eq:D_5d_matrix}) with the
replacement $D_- \to -1$ and setting $\hat A(m) =0$, we see that
\begin{eqnarray}
\left[ D^{(5)}_{DW}(1) - D^{(5)}_{DW}(m) \right]_{ij} = (1-m) \left[
 P_- \delta_{iL_s} \delta_{j1} + P_+ \delta_{i1}\delta_{jL_s} \right] ~.
\end{eqnarray}
Therefore we obtain
\begin{eqnarray}
D_{tov}^{-1}(m;H_T) - 1 = (1-m) \left\{ {\cal P}^{-1} [ D^{(5)}_{DW}(m) ]^{-1}
 {\cal J} {\cal P} \right\}_{11} ~,
\end{eqnarray}
where ${\cal J}_{ij} = \delta_{i,L_s+1-j}$ is the inversion operator of the
fifth direction. Now from (\ref{eq:light}) we see that the physical
fermion degrees are given in terms of the domain wall boundary fermions
as
\begin{eqnarray}
q = ({\cal P}^{-1} \Psi)_1 ~, \qquad\qquad
\bar q = (\bar \Psi {\cal J} {\cal P})_1 ~.
\end{eqnarray}
Hence we find
\begin{eqnarray}
\langle q \bar q \rangle = \left\{ {\cal P}^{-1} [ D^{(5)}_{DW}(m) ]^{-1} {\cal J}
 {\cal P} \right\}_{11} = \frac{1}{1-m} \left[ D_{tov}^{-1}(m;H_T) - 1 \right] ~.
\label{eq:q_prop}
\end{eqnarray}
The usual domain wall physical fermion propagator automatically contains
both the subtraction and multiplicative normalization of the overlap
fermion propagator.

The hermitian conjugate of the 5-d operator $D^{(5)}_{DW'}$,
eq.~(\ref{eq:D_5d_matrix}), of Bori\c{c}i's variant is different than that
of the usual DWF operator, $D^{(5)}_{DW}$, without projection.
In particular it does not have the (generalized) $\gamma_5$ hermiticity,
$D^{(5)\dagger}_{DW} = \gamma_5 {\cal J} D^{(5)}_{DW} \gamma_5 {\cal J}$,
with ${\cal J}$ the inversion operator of the fifth direction.
Instead we find
\begin{eqnarray}
D^{(5)\dagger}_{DW'} = \begin{pmatrix}
D_+^\dagger - P_-\hat{A}^\dagger & P_+D_-^\dagger & 0 & 0&\cdots & 0 & 0 &
-m P_- D_-^\dagger \cr
P_- D_-^\dagger & D_+^\dagger & P_+ D_-^\dagger & 0&\cdots & 0 & 0 & 0 \cr
0 & P_- D_-^\dagger & D_+^\dagger & P_+ D_-^\dagger &\cdots & 0 & 0 & 0 \cr
\vdots & \vdots & \vdots & \vdots & \vdots & \vdots & \vdots & \vdots \cr
0 & 0 & 0 & 0 & \cdots &  P_- D_-^\dagger & D_+^\dagger & P_+ D_-^\dagger \cr
-m P_+ D_-^\dagger - P_+\hat{A}^\dagger & 0 & 0 & 0 & \cdots & 0 & 
P_- D_-^\dagger & D_+^\dagger \cr
\end{pmatrix}
\label{eq:Ddag_5d_matrix}
\end{eqnarray}
where
\begin{eqnarray}
D^\dagger_\pm = a_5 D_w^\dagger(-M) \pm 1,\qquad
\hat{A}^\dagger(m) = (1-m)\sum_i f_i\hat{P}_i \gamma_5\ .
\end{eqnarray}

\subsection{Eigenvalues of the pseudo-fermion operators}
\label{sec:pseudo}

Here, we want to investigate a little more the properties of the
pseudo-fermions, needed to cancel the bulk contributions in the 5-d
domain wall fermion approaches.

For both the standard domain wall fermion action (\ref{eq:S_DW}) and
for Bori\c{c}i's variant (\ref{eq:S_B}) we find for the pseudo-fermion
matrix, including the boundary conditions and recalling that the
projection operator $\hat A(1)$ vanishes,
\begin{eqnarray}
D^{(5)}(1) D^{(5)\dagger}(1) = \begin{pmatrix}
X & Y & 0 & 0 &\cdots & 0 & 0 & -Y \cr
Y & X & Y & 0 &\cdots & 0 & 0 & 0 \cr
0 & Y & X & Y &\cdots & 0 & 0 & 0 \cr
\vdots & \vdots & \vdots & \vdots & \vdots & \vdots & \vdots & \vdots \cr
0 & 0 & 0 & 0 & \cdots & Y & X & Y \cr
-Y & 0 & 0 & 0 & \cdots & 0 & Y & X \cr
\end{pmatrix}
\label{eq:D_Ddag_5d}
\end{eqnarray}
with $X$ and $Y$ 4-d hermitian matrices,
\begin{eqnarray}
Y &=& - P_+ D_+ - D_+^\dagger P_- = -\frac{1}{2} \left(a_5 D_w +
 a_5 D_w^\dagger + 2 \right) \nonumber \\
X &=& D_+ D_+^\dagger + 1 = a_5^2 D_w D_w^\dagger - 2 Y
\label{eq:X_Y_DW}
\end{eqnarray}
for the standard domain wall action, and
\begin{eqnarray}
Y &=& D_+ P_+ D_-^\dagger + D_- P_- D_+^\dagger = a_5^2 D_w D_w^\dagger - 1
 \nonumber \\
X &=& D_+ D_+^\dagger + D_- D_-^\dagger = 2 a_5^2 D_w D_w^\dagger + 2 =
 4 a_5^2 D_w D_w^\dagger - 2 Y
\label{eq:X_Y_B}
\end{eqnarray}
for Bori\c{c}i's variant. Similar relations can be found in
Ref.~\cite{Shamir_a5}. 

Now, let $S$ be the shift (or translation) operator in the 5-th direction
with anti-periodic boundary condition:
\begin{eqnarray}
S = \begin{pmatrix}
0 & 1 & 0 & \dots & 0 & 0 \cr
0 & 0 & 1 & \dots & 0 & 0 \cr
\vdots & \vdots & \vdots & \vdots & \vdots & \vdots \cr
0 & 0 & 0 & \dots & 0 & 1 \cr
-1 & 0 & 0 & \dots & 0 & 0 \cr \end{pmatrix}
\label{eq:shift}
\end{eqnarray}
It is easy to see that $[S, D^{(5)}(1) D^{(5)\dagger}(1) ] = 0$, and so
$S$ and $D^{(5)}(1) D^{(5)\dagger}(1)$ can be diagonalized simultaneously.
The eigenvalues of $S$ are $\exp\{i\pi(2k+1)/L_s\}$ for $k=0,\dots,L_s-1$,
because of the anti-periodic boundary conditions. The corresponding
eigenvectors are
\begin{eqnarray}
\label{eq:ev_dwpf}
&& w(k)^T = \\
&& (v_4, v_4 \exp\{i\pi(2k+1)/L_s\}, v_4 \exp\{i\pi 2(2k+1)/L_s\}, \dots,
v_4 \exp\{i\pi(L_s-1)(2k+1)/L_s\} )^T \nonumber
\end{eqnarray}
with $v_4$ some 4-d vector. This also has to be an eigenvector of
$D^{(5)}(1) D^{(5)\dagger}(1)$. From (\ref{eq:D_Ddag_5d}), and using
(\ref{eq:X_Y_DW}) and (\ref{eq:X_Y_B}) we obtain the eigenvalue
equation for $v_4$
\begin{eqnarray}
\left\{ J a_5^2 D_w D_w^\dagger - 2 Y \left[ 1 - \cos \left( \frac{\pi}{L_s}
(2k+1) \right) \right] \right\} v_4 = \lambda(k) v_4
\label{eq:4d_pf_ev}
\end{eqnarray}
Here $J=1$ for domain wall fermions, and $J=4$ for Bori\c{c}i's variant,
and we have indicated the dependence of $\lambda$ on $k$, the momentum
in the 5-th direction. Since the left-hand-side of (\ref{eq:4d_pf_ev})
remains unchanged under $k \rightarrow L_s-k-1$ we conclude that
$\lambda(L_s-k-1) = \lambda(k)$. Therefore, the eigenvalues of
$D^{(5)}(1) D^{(5)\dagger}(1)$ are all (at least) doubly degenerate.

Furthermore, for large $L_s$ and small $k$ the pseudo-fermion domain wall
eigenvalues closely track the Wilson eigenvalues, the eigenvalues of
$D_w D_w^\dagger$, since then $2 Y \left[ 1 - \cos \left( \frac{\pi}{L_s}
(2k+1) \right) \right] \approx Y \pi^2 (2k+1)^2/L_s^2$ in (\ref{eq:4d_pf_ev})
is only a small perturbation. In particular, at zero crossings of
the Wilson Dirac operator, a pair of pseudo-fermion domain wall eigenvalues
will go to zero for $L_s \to \infty$.

{}From the domain wall Dirac operator $D^{(5)}_{DW}(m)$ without the
projection matrix $\hat A(m)$, we can make a hermitian version,
$H^{(5)}_{DW}(m) = D^{(5)}_{DW}(m) {\cal J} \gamma_5$, with
${\cal J}_{ij} = \delta_{i,L_s+1-j}$ the inversion operator of the fifth
direction. From eq.~(\ref{eq:D_5d_matrix}) with $D_- \to -1$ and
$\hat A =0$ it becomes
\begin{eqnarray}
H^{(5)}_{DW}(m) = \begin{pmatrix}
m P_+ & 0 & 0 & 0 & \cdots & 0 & P_- & D_+ \gamma_5 \cr
0 & 0 & 0 & 0 & \cdots & P_- & D_+ \gamma_5 & -P_+ \cr
0 & 0 & 0 & 0 & \cdots & D_+ \gamma_5 & -P_+ & 0 \cr
\vdots & \vdots & \vdots & \vdots & \vdots & \vdots & \vdots & \vdots \cr
P_- & D_+ \gamma_5 & -P_+ & 0 & \cdots & 0 & 0 & 0 \cr
D_+ \gamma_5 & -P_+ & 0 & 0 & \cdots & 0 & 0 & -mP_- \cr
\end{pmatrix}
\label{eq:H_5d_DW}
\end{eqnarray}
Since $D^{(5)}_{DW}(m) D^{(5)\dagger}_{DW}(m) = [H^{(5)}_{DW}(m)]^2$, the
vectors $w(k)$ and $w(L_s-k-1)$ of eq.~(\ref{eq:ev_dwpf}) are degenerate
eigenvectors of $[H^{(5)}_{DW}(1)]^2$. To get the eigenvalues of
$H^{(5)}_{DW}(1)$ we need to compute the $2 \times 2$ matrix $\langle w(k_1) |
H^{(5)}_{DW}(1) | w(k_2) \rangle$ with $k_1, k_2 = k, L_s-k-1$. We find
\begin{eqnarray}
&& \langle w(k_1) | H^{(5)}_{DW}(1) | w(k_2) \rangle = \nonumber \\
&& \sum_{j=1}^{L_s}
 {\rm e}^{ -\frac{i\pi}{L_s}(2k_1+1)(j-1)} {\rm e}^{ \frac{i\pi}{L_s}(2k_2+1)(L_s-j)}
 \langle v_4 | P_- {\rm e}^{ -\frac{i\pi}{L_s}(2k_2+1)} + D_+ \gamma_5
 - P_+ {\rm e}^{ \frac{i\pi}{L_s}(2k_2+1)} | v_4 \rangle 
\end{eqnarray}
The sum over $j$ gives the constraint $k_2 = L_s-k_1-1$, and we thus find
the $2 \times 2$ matrix to be of the form
\begin{eqnarray}
\begin{pmatrix} 0 & b \cr b^* & 0 \cr \end{pmatrix} \nonumber
\end{eqnarray}
and therefore the eigenvalues of the hermitian pseudo-fermion domain wall
operator come in $\pm$ pairs.

\subsection{Preconditioning}
\label{sec:precond}

Even-odd preconditioning is a common technique to reduce the condition
number of a matrix~\cite{precond} and can be implemented
in both versions of the domain wall actions discussed above. With
projection, the matrices have a more complicated form reminiscent of
preconditioning for the clover fermion action.

Write the matrix $D^{(5)}$ as a two by two block matrix
\begin{eqnarray}
D^{(5)} = \begin{pmatrix} {\cal A}_{EE} & {\cal B}_{EO} \\
  {\cal B}_{OE} & {\cal A}_{OO} \end{pmatrix}
\end{eqnarray}
where the upper case characters $E$ and $O$ label even and odd
(checkerboard) sites, respectively, in five dimensions. The matrix
$D^{(5)}$ can be brought into an even-odd block diagonal form using
the lower and upper block triangular matrices
\begin{eqnarray}
L = \begin{pmatrix} I_{EE} & 0 \\ 
{\cal B}_{OE}{\cal A}_{EE}^{-1} & I_{OO} \end{pmatrix}
\qquad
U = \begin{pmatrix} {\cal A}_{EE} & {\cal B}_{EO}\\
0 & I_{OO} \end{pmatrix}
\end{eqnarray}
with the transformation
\begin{eqnarray}
{\tilde D}^{(5)} = L^{-1} D^{(5)} U^{-1} = 
\begin{pmatrix} I_{EE} & 0 \\ 0 &
{\cal A}_{OO} - {\cal B}_{OE}{\cal A}_{EE}^{-1}{\cal B}_{EO}
\end{pmatrix}.
\end{eqnarray}
This decomposition always exists as long as ${\cal A}_{EE}$ is
non-singular. 

The structure of ${\cal A}_{EE}$ is deduced from the structure of
$D_w(M)$ and the projection matrix ${\hat A}$ that occur in both the
standard domain wall and Bori\c{c}i variants, $D^{(5)}_{DW}$ and
$D^{(5)}_{DW'}$, respectively. In Bori\c{c}i's variant the nearest neighbor
coupling in the fifth direction contains also hoppings in the spatial
directions leading to a complicated even-odd structure. This case will not
be discussed further here.  For preconditioning of the standard domain wall
action with projection, after a rescaling of the fermion fields to bring
$D^{(5)}_{DW}$ into the ``kappa''-form, {\it i.e.}, having 1 in the diagonal
except for the projection matrix $\hat A$, we have
\begin{eqnarray}
{\cal A}_{EE} = \begin{pmatrix}
1 - \hat{A}_{ee} P_- & 0 & \cdots & 0 & - \hat{A}_{eo} P_+ \cr
0 & 1 & \cdots & 0 & 0 \cr
\vdots & \vdots & \vdots & \vdots & \vdots \cr
0 & 0 & \cdots & 1 & 0 \cr
0 & 0 & \cdots & 0 & 1 \cr
\end{pmatrix} ~.
\end{eqnarray}
Here $\hat{A}$ is rescaled by $2 \kappa$, with
$\kappa=1/(2(a_5(4+M)+1))$, and $e$ and $o$ refer to the even and
odd sub-lattice in the 4-d checker-boarding. 

For even-odd preconditioning, we need
\begin{eqnarray}
{\cal A}^{-1}_{EE} = \begin{pmatrix}
(1 - \hat{A}_{ee} P_-)^{-1} & 0 & \cdots & 0 & (1 - \hat{A}_{ee} P_-)^{-1}
 \hat{A}_{eo} P_+ \cr
0 & 1 & \cdots & 0 & 0 \cr
\vdots & \vdots & \vdots & \vdots & \vdots \cr
0 & 0 & \cdots & 1 & 0 \cr
0 & 0 & \cdots & 0 & 1 \cr
\end{pmatrix} ~.
\label{eq:precond}
\end{eqnarray}
Therefore, for every application of the projected domain wall Dirac
operator, the inversion $(1 - \hat{A}_{ee} P_-)^{-1} \phi$ for some
4-d even sub-lattice fermion $\phi$ has to be performed once --- the two
occurrences in ${\cal A}^{-1}_{EE}$ can be combined into the
application on one 4-d even sub-lattice fermion field.  In the limit of
the fermion mass $m\rightarrow 1$, the matrices ${\cal A}^{-1}_{EE}$
and ${\cal A}_{OO}$ become the unit matrix. For small $m$, the matrix
$(1 - \hat{A}_{ee} P_-)$ should be well conditioned, and the cost of
inversion is the number of iterations involving applications of $D_w$
on half of the four volume. This ``inner'' iteration cost is a
multiplicative overhead on the cost of inverting ${\tilde D^{(5)}}$.

\subsection{Dynamical fermions}
\label{sec:dynamical}

Implementing an HMC algorithm for dynamical domain wall fermion simulations
is a straightforward modification of dynamical Wilson fermion simulation
code. Implementing projection does not cause any fundamental difficulties.
What is needed, in addition to the usual case, for the computation of the
fermion contribution to the force is $\partial \hat A(m;H(U)) / \partial U$,
where we indicated the implicit dependence on $U$ of the projection operator
$\hat A$. Using the chain rule, what we need is
\begin{eqnarray}
\frac{\partial \lambda_i}{\partial U} \qquad {\rm and} \qquad
\frac{\partial \hat P_i}{\partial U} = \frac{\partial v_i}{\partial U}
 v^\dagger_i + v_i \frac{\partial v^\dagger_i}{\partial U} ~,
\end{eqnarray}
for $H(U) v_i = \lambda_i v_i$ with $v^\dagger_i v_i = 1$. We easily find
\begin{eqnarray}
\frac{\partial \lambda_i}{\partial U} = v^\dagger_i
 \frac{\partial H(U)}{\partial U} v_i ~.
\end{eqnarray}
and
\begin{eqnarray}
\frac{\partial v_i}{\partial U} = \left( H - \lambda_i \right)^{-1}
 \left[ v^\dagger_i \frac{\partial H(U)}{\partial U} v_i -
 \frac{\partial H(U)}{\partial U} \right] v_i ~.
\end{eqnarray}
Therefore, a different 4-d inversion is needed for every projected
eigenvector $v_i$. For the standard domain wall fermion action, where
$H = H_T$, additional 4-d inversions are needed to compute
$\frac{\partial H_T(U)}{\partial U}$.

We note that the modifications involving projection for the force term
may not be necessary in practice. The unprojected action can be used
to construct the guiding Hamiltonian for HMC with the projected action
used for the computation of the initial and final energies in the
Metropolis accept/reject step. A critical factor in the choice will be
the acceptance rate which will suffer if the guiding Hamiltonian is
not accurate enough without projection.

\section{Results}
\label{sec:results}

\subsection{Smooth gauge field}

We start our analysis by first considering the case of spectral flow
on a smooth SU(2) instanton configuration. We use the instanton
construction from ref.~\cite{EHN_smooth}, namely a single $8^4$
instanton with $\rho=1.5$ and anti-periodic boundary conditions. 
In this section we will mainly focus on the standard domain wall
operator. Similar results are found for the Bori\c{c}i variant.

What we are interested in is first determining how various domain-wall
like actions reproduce topology on the lattice. Without projection, we
note that the index of the massless four dimensional (4-d) operator is
\begin{eqnarray}
Q = {\rm Tr}(\gamma_5 D_{tov}) = 
  \frac{1}{2}{\rm Tr}\left(\varepsilon_{L_s/2}(a_5 H(-M))\right)
\end{eqnarray}
with the auxiliary Hamiltonian $H$ given by either $H_T$ or $H_w$ ---
the two cases under consideration. For infinite $L_s$,
$\varepsilon_{L_s/2}$ becomes $\epsilon$ and $Q$ is a measure of the
discrepancy of positive and negative states of $H$. A simple way to
determine $Q$ is to start from some $M$ where we know that $Q=0$ and
increase it to the desired value while counting the change of the
number of positive states. We emphasize that this procedure works as long
as the spectral flow is smooth --- namely, as long as the eigenvalues of
$H$ are well behaved. While $H_w$ is always well behaved, we note that
this is not the case for the domain wall $H_T(-M)$ for $M > 2$. In
particular, the spectral flow method will {\em fail} for $M$ in the
``doubler'' region where there are multiple species of chiral fermions.
We will address this point in more detail later.

In Fig.~\ref{fig:dwf_flow} we show the spectral flow of the lowest 10
eigenvalues of the hermitian Wilson-Dirac operator $H_w(-M)$, the
lowest 10 eigenvalues of the hermitian domain wall Dirac operator
${\cal J}\gamma_5 D^{(5)}_{DW}(0)$ with $a_5=1$ for various fifth
dimensional extent values $L_s$, and the 10 lowest eigenvalues of the
hermitian overlap-Dirac operator (with projection, {\it i.e.} with
chiral symmetry), all as function of the ``domain wall height $M$''.
We recall from eq.~(\ref{eq:op5d4d}) 
\begin{eqnarray}
D_{ov}(0) = \left\{ {\cal P}^{-1} \left[ D^{(5)}(1) \right]^{-1}
 D^{(5)}(0) {\cal P} \right\}_{11}
\label{eq:op5d4d_again}
\end{eqnarray}
that when $D_{ov}(0)$ has zero-eigenvalues we can expect $D^{(5)}(0)$
to have zero-eigenvalues as long as
$D^{(5)}(1)$ is not singular. We see in Fig.~\ref{fig:dwf_flow} that there
are exact zeros of $D_{ov}(0)$ for $M$ beyond the crossing, and the
spectrum changes discontinuously at the crossing~\cite{EHN_practical}.
For finite $L_s$, the $D^{(5)}_{DW}(0)$ eigenvalues decrease quickly in
$L_s$. 
For $M$ beyond the crossing they quickly go to
zero and in fact appear exponential in $L_s$, but near the crossing, when
the eigenvalues of $H_w$ are small, the decrease is slowed.

This slow rate of convergence in $L_s$, when the eigenvalues of $H_w$
are small, is a prime source of difficulty in recent numerical
calculations involving domain wall fermions. The reason for
the slow rate of convergence is clear when the form of the 4-d
operator is examined,
\begin{eqnarray}
D_{tov}(m) = 
\frac{1}{2} \Bigl[1+m + (1-m) \gamma_5 \varepsilon_{L_s/2}(a_5 H) \Bigr]~.
\label{eq:d_tov_again}
\end{eqnarray}
When $\varepsilon_{L_s/2}(a_5 H)$ deviates from one, for example for the
smallest and largest eigenvalues of $H$, there is violation
of chiral symmetry. Fig.~\ref{fig:dwf_proj_flow} shows the domain wall
spectral flow when the $5$ smallest (closest to zero) eigenvectors of $H_T$
are projected out as prescribed by
eqs.~(\ref{eq:hat_A_dwf}),(\ref{eq:g_i}) and
(\ref{eq:D_ov_proj}). With projection turned on, $D^{(5)}_{DW}(0)$ no
longer has (generalized) $\gamma_5$ hermiticity, so the eigenvalues of
$+\sqrt{(D^{(5)}_{DW}(0))^{\dagger} D^{(5)}_{DW}(0)}$ are
determined. We see that just beyond the crossing there is a
discontinuous drop of the eigenvalues to zero. In fact fairly small
eigenvalues are seen even for $L_s = 4$. Also shown are the lowest
eigenvalues of $H_T$. We see they agree quite well with the
eigenvalues of $H_w$ up to a factor of two as predicted by
eq.~(\ref{eq:H_T}).

All computations of eigenvalues and eigenvectors where done with the
Ritz method~\cite{Ritz}. For computing eigenvectors of $H_T(-M)$, in
each Ritz iteration, the application of
$H_T(-M)^\dagger H_T(-M)$ on a vector is needed. From
eq.~(\ref{eq:H_T}) for $H_T$, the expression for $H_T^\dagger H_T$ can
be combined into
\begin{eqnarray}
H_T^\dagger H_T = \gamma_5 D_w \frac{1}{(2 + a_5 D_w^\dagger)(2+a_5
D_w)} \gamma_5 D_w~,
\end{eqnarray}
so a single inversion of a hermitian positive definite operator
is needed for each Ritz iteration and Conjugate Gradient for Normal
Equations (CGNE) was used. Typically 30 to 60 iterations of CGNE were
needed for each inversion to achieve an accuracy of $10^{-7}$.

It should come as no surprise that zero eigenvalues of the domain wall
operator can be obtained when projection is turned on since this is
the same mechanism by which they are obtained for the overlap-Dirac
operator~\cite{EHN_practical}. What was key to the latter case was
enough projection coupled with a judicious choice of the fifth
dimension spacing $a_5$. Since $a_5$ only serves as a multiplicative
scaling of $H_w$, $a_5$ can be chosen so that the lowest unprojected
and highest eigenvalues of $H_w$ lie within the range of approximation
of $\epsilon$. For the overlap-Dirac operator in
Ref.~\cite{EHN_practical}, an optimal rational approximation was
chosen; however, this merely increased the useful range of the
approximation. When the condition is met for a valid approximation of
$\epsilon$, there are no other cutoff effects associated with having a
finite $a_5$. For example, observables of the effective 4-d
theory will not have slightly different 4-d
lattice spacing dependencies for slightly different choices of $a_5$.
However, for the conventional domain wall action, $H_T$ intrinsically
depends on $a_5$ as seen in eq.~(\ref{eq:H_T}). Therefore, even for infinite
extent in the fifth direction, where $\varepsilon(a_5 H_T)$ converges
to $\epsilon(a_5 H_T)=\epsilon(H_T)$,
choosing different $a_5$ results in different 4-d
lattice spacing dependence for physical observables. This
fact limits the usefulness of adjusting $a_5$ with the goal of imposing
arbitrarily precise chiral symmetry for (small) finite fifth dimension
as will be shown later.

In Fig.~\ref{fig:compare_dwftrf} we examine more closely the smallest
eigenvalue of the domain wall operator and the deviation of
$\varepsilon(a_5 H_T)$ from one for $a_5 = 1$.
We see that without projection the eigenvalues decrease
approximately exponentially in $L_s$ with a slow rate varying with $M$,
while with projection they drop to about $10^{-4}$ after the crossing.
This is the accuracy to which the eigenvalues were computed in single
precision and is ``zero'' here. For increasing $M$, the zero
eigenvalues slowly increase. 

Some clue to the slow increase can be seen from the right panel in
Fig.~\ref{fig:compare_dwftrf} which shows the deviation from unity of
$\varepsilon(\lambda_i)$ where $\lambda_i$ is the first, fifth and
largest eigenvalue of $H_T$. The deviation for the fifth and the
largest eigenvalue gives a measure of the accuracy of the
approximation since the first five eigenvalues are projected out of
$H_T$. The deviation of the first eigenvalue indicates  how much the
standard unprojected case deviates. For $L_s=4$ the
deviation of the largest eigenvalue is clearly visible, and for
$L_s\ge 8$ the deviation is off the bottom of the graph until $M$
approaches $2$. For $L_s=8$ a typical deviation of the fifth eigenvalue
is about 1\% while nonetheless a domain wall eigenvalue of $10^{-4}$
is obtained.  However, as $M$ is increased, the largest eigenvalue is
not well approximated and the domain wall eigenvalue deviates away from
zero due to the resulting chiral symmetry breaking.

\begin{figure}
\centerline{{\setlength{\epsfxsize}{6in}\epsfbox[0 0 576 576]{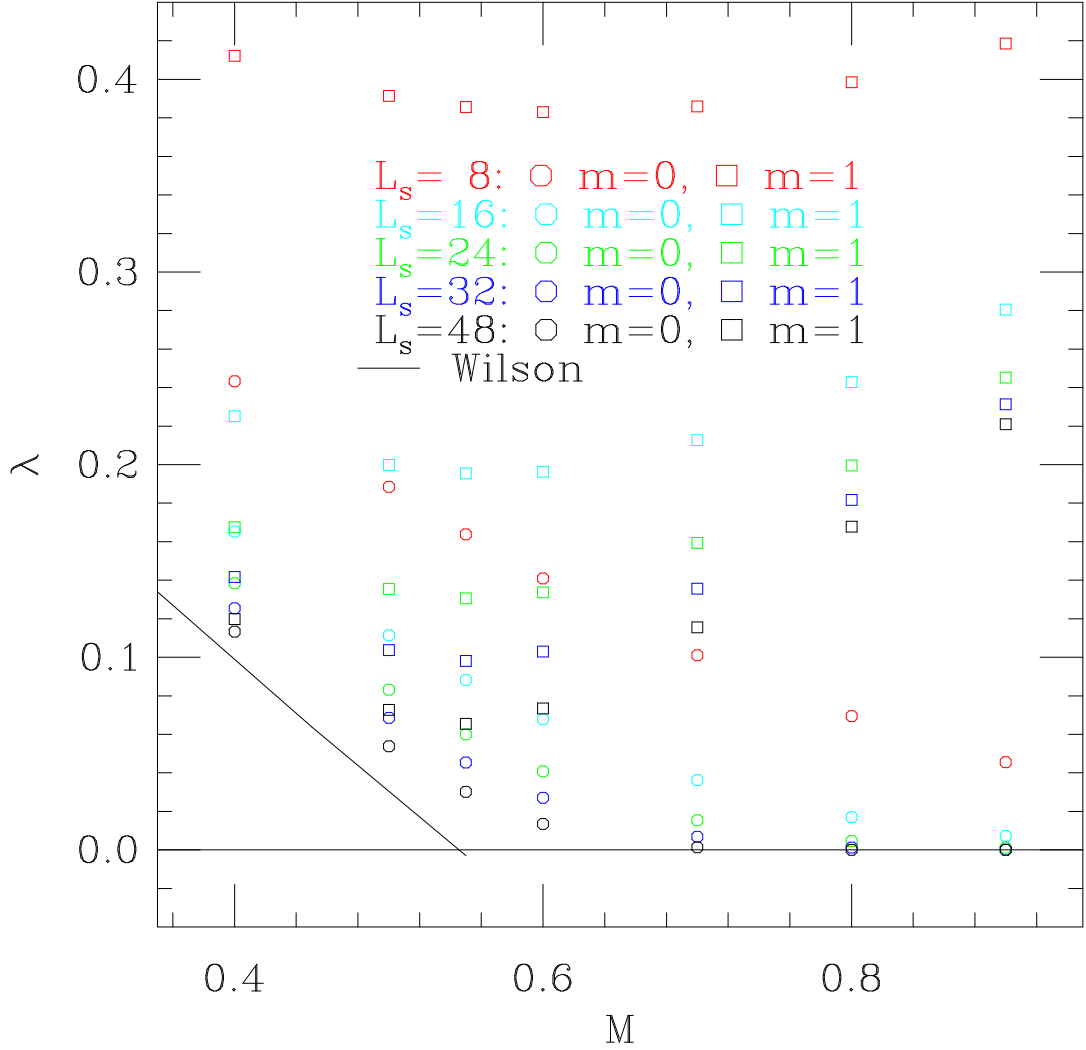}}}
\caption{Spectral flow of the lowest pseudo-fermion eigenvalue ($m=1$)
and lowest domain wall eigenvalue ($m=0$, with no projection)
for various $L_s$ on a single $8^4$ instanton background. The minimum
of the lowest pseudo-fermion eigenvalues is at the Wilson crossing and
decreases with increasing $L_s$.}
\label{fig:dwf_low_pf}
\end{figure}

We will conclude this portion of the analysis with a final look at the
role of the pseudo-fermion term in eq.~(\ref{eq:op5d4d}) (repeated
again in eq.~(\ref{eq:op5d4d_again})). It was stated that zero
eigenvalues of the 5-d $D^{(5)}(0)$ are seen when the
4-d $D_{ov}(0)$ has zero eigenvalues. However, as shown
in Sec.~\ref{sec:pseudo}, the pseudo-fermion term has a zero
eigenvalue in the infinite fifth dimensional extent limit at the zero
crossings of $H_w$. We show in Fig.~\ref{fig:dwf_low_pf} the lowest
eigenvalue 
of the hermitian version of the unprojected $D^{(5)}(0)$ and the
pseudo-fermion operator $D^{(5)}(1)$ (recall that the pseudo-fermion
operator is unaffected by projection). The pseudo-fermion eigenvalues
track the regular eigenvalues up to the crossing, then move away from zero
again. As $L_s$ increases, the pseudo-fermion eigenvalues appear to go
exponentially to zero. We see then by studying
eq.~(\ref{eq:op5d4d_again}) that the pseudo-fermions cancel to some
extent in the ratio $(D^{(5)}(1))^{-1} D^{(5)}(0)$. 

For the resultant
unprojected 4-d operator $D_{tov}(0)$ (finite $L_s$), the
eigenvalues are larger than the corresponding 5-d ones
and the 4-d eigenvalues decrease across a broad region in the mass
$M$. These eigenvalues are the actual ones occurring in 4-d
observables like the pion propagator and the chiral condensate. More to
the point, the eigenvalues of the 5-d operator $D^{(5)}(m)$
are in general not directly physically relevant to 4-d
observables. Stated differently, the eigenvectors or some piece of the
eigenvectors of the $D^{(5)}(m)$ are not eigenvectors of $D_{tov}(m)$
because of the basis changing matrices ${\cal P}$ and the projection
onto the first 5-d slice. However,
eq.~(\ref{eq:op5d4d_again}) gives us the connection between the 5-d and
4-d matrices and in fact provides a new algorithm to
compute the relevant 4-d eigenvalues from only the 5-d
operators. It has some attractive features for practical
use because of its simplicity and direct connection, but it is not really
efficient for the reason that $D^{(5)}(1)$ has near zero eigenvalues
at $H_w$ zero crossings and hence has bad condition numbers, at least
in the large $L_s$ limit.

We proceed to another test of chiral symmetry given by the generalized
Gellmann-Oakes-Renner relation (called GMOR by many authors). In the
form used here and with our normalization
conventions~\cite{EHN_practical}, it states that
\begin{eqnarray}
m \;\langle b| (\gamma_5 {\tilde D}^{-1}_{ov}(m))^2 |b\rangle =
\langle b|{\tilde D}^{-1}_{ov}(m)|b\rangle\qquad 
\forall\quad \gamma_5 |b\rangle = \pm |b\rangle ~,
\label{eq:GMOR}
\end{eqnarray}
with the physical (subtracted, see eq.~(\ref{eq:4d5d_prop})) fermion
propagator
\begin{eqnarray}
{\tilde D_{tov}}^{-1}(m) =\frac{1}{1-m} \left[D_{tov}^{-1}(m) - 1 \right]
\label{eq:phys_prop}
\end{eqnarray}
and $L_s$ taken to the infinite fifth dimensional limit.
Averaging over several (chiral) Gaussian random vectors $|b\rangle$,
eq.~(\ref{eq:GMOR}) becomes a stochastic estimate for $m \chi_\pi =
\langle \bar \psi \psi \rangle$, which is the familiar Gellmann-Oakes-Renner
relation.
Eq.~(\ref{eq:GMOR}) holds configuration by configuration for any
chiral state $|b\rangle$. For finite fifth dimensional extent $L_s$, the
relation is broken and a useful measure is the ratio
\begin{eqnarray}
R = m \frac{\langle b|(\gamma_5 {\tilde D}^{-1}_{tov}(m))^2|b\rangle}{
\langle b|{\tilde D}^{-1}_{tov}(m)|b\rangle}~.
\label{eq:GMOR_ratio}
\end{eqnarray}
This ratio has been studied for domain wall
fermions~\cite{Columbia_GMOR} where in quenched SU(3) configurations at
$\beta=5.85$ and $5.7$ significant deviations of the ratio from one are
seen for fermion masses on the order $m=0.01$ and smaller. We argue that
the larger violations seen at $\beta=5.7$ are due to the larger ensemble
averaged density of zero eigenvalues of $H_w(-M)$ at $\beta=5.7$
compared to $\beta=5.85$~\cite{EHN_rho0}.

We show in Fig.~\ref{fig:smooth_GMOR} a plot of the ratio $R$ for the
domain wall operator as a function of the fermion mass at a fixed
$M=0.5$ and $a_5=1$ with the instanton background used in
Fig.~\ref{fig:dwf_flow}. The mass $M=0.5$ is just before the zero
crossing of $H_w(-M)$. Without projection, the deviation is large even for
$L_s=32$. With $5$ eigenvalues projected, the deviation from unity
is negligible for $L_s=16$. For $L_s=8$, the violation is large with
$5$ eigenvectors projected out and is not noticeably changed with $20$
eigenvectors projected out. This is explained by noticing in the right
panel of Fig.~\ref{fig:compare_dwftrf} that for $L_s=8$ the fifth
eigenvalue shows a deviation of $1-|\varepsilon_4(\lambda_5)|\approx
0.01$. What is not shown is that this value changes little at the 20th
eigenvalue. What we are seeing is that by the 20th eigenvalue we are
entering into a relatively dense band of eigenvalues of $H_T$ and they
change slowly with increasing eigenvalue. Therefore, little is gained
by projecting out more eigenvectors. By going to $L_s=16$,
$1-|\varepsilon_8(\lambda_5)|\approx 5\times 10^{-5}$ is much reduced
compared to $L_s=8$, and hence all the volume modes contributing to
$R$ have deviation less (or much less) than the worst case resulting
in a small overall deviation of $R$ from unity. Therefore, projection
is quite effective at restoring chiral symmetry, and we see that the
GMOR test given by eq.~(\ref{eq:GMOR_ratio}) is a quite sensitive measure of
chiral symmetry breaking since it detected a deviation in the
$\epsilon$ approximation on order of 1\%.

We finish up the discussion of the smooth field case by considering
the case of $M>2$ (the doubler region). We continue the discussion
initiated in Ref.~\cite{Shamir_a5} on the difference of the overlap
Dirac operator $D_{ov}(m;H_w)$ and the 4-d version of the standard
domain wall operator $D_{ov}(m;H_T)$. 

We consider the free propagator with momentum near one of the corners
of the Brillouin zone, {\it i.e.,} with ${\overline p_\mu} = \sin(p_\mu) \ll 1$
for all $\mu$. Set $M(p) = M - B(p) = M - \sum_\mu 2 \sin^2(p_\mu/2)
= M - 2n + {\cal O}({\overline p_\mu}^2)$, where $n$ is the number of
momentum components near $\pi$. Then $D_w(-M) = i \sum_\mu \gamma_\mu
{\overline p_\mu} - M(p)$ and we find for the inverse free overlap
propagator, up to terms of higher order in ${\overline p_\mu}$ and $m$,
\begin{eqnarray}
\tilde D^{-1}_{ov}(p,m) = \begin{cases}
2 M(p) \times \frac{-i \sum_\mu \gamma_\mu {\overline p_\mu} + 2 m M(p)}{
\sum_\mu {\overline p_\mu}^2 + (2 m M(p))^2} & \text{for $M(p) > 0$} \\
\frac{-i \sum_\mu \gamma_\mu {\overline p_\mu}}{2 |M(p)|}
& \text{for $M(p) < 0$}~. \end{cases}
\label{eq:free_ov}
\end{eqnarray}
This corresponds to a light free fermion propagator only for $M(p) \approx
M - 2n > 0$. For $2 < M < 4$, for example, the origin and the corners
of the Brillouin zone with one momentum component close to $\pi$ give
light fermions, while for $0 < M < 2$ only the origin gives a light
fermion.

For free domain wall fermions, recalling the form of $H_T$ from
eq.~(\ref{eq:H_T}),
\begin{eqnarray}
H_T(-M) = \gamma_5 D_w(-M) \frac{1}{2 + a_5 D_w(-M)}
 = \gamma_5 D_w(-M) \frac{1}{a_5 D_w(\frac{2}{a_5} -M)}~.
\label{eq:H_T2}
\end{eqnarray}
we find
\begin{eqnarray}
D_T(-M) = \gamma_5 H_T(-M) = C(p) \left[ i \sum_\mu \gamma_\mu
 {\overline p_\mu} - \frac{1}{2} M(p) \left( 2 - a_5 M(p) \right) \right]
\end{eqnarray}
with $C(p)$ positive, and hence unimportant in $\epsilon(H_T)$. We see that
for free domain wall fermions $\frac{1}{2} M(p) \left( 2 - a_5 M(p) \right)$
plays the role of $M(p)$ for overlap fermions. Hence, from
eq.~(\ref{eq:free_ov}) we see that we need $M(p) \left( 2 - a_5 M(p) \right)
> 0$ to have a light domain wall fermion. These are exactly the conditions
for a normalizable domain wall zero mode~\cite{Shamir_a5}, $-1 < 1 - a_5 M(p)
< 1$. The first inequality here is an additional condition for domain
wall fermions, not present for overlap fermions. For $2 < M < 4$ it
excludes the origin of the Brillouin zone from giving a light fermion,
as it did for overlap fermions.

Now consider the interacting case.  We know that zero crossings of
$H_w(-M)=0$ also correspond to zero crossings of the domain wall
$H_T(-M)$ for all $a_5$.  However, eq.~(\ref{eq:H_T2}) suggests that
$H_T(-M)$ can have poles.
Zero eigenvalues of $H_w(-M)$ at some $M_0$ are also zero eigenvalues
of $D_w(-M_0)$. From (\ref{eq:H_T2}) we see that to each such zero
eigenvalue is associated a pole of $H_T(-M)$ at $M=M_0 + 2/a_5$. In other
words, all zeros below $M=2$ are replicated as poles at $M + 2/a_5$.

For the single instanton background shown in Fig.~\ref{fig:dwf_flow},
there is one downward crossing in the Wilson spectral flow at about
$M=0.55$. After $M=0.55$, the domain wall and overlap Dirac operators
have index $-1$ and hence a single zero mode. For this one downward
crossing, there are 4 corresponding upward crossings at 
$M=2.17$ and $M=2.30$. After this last crossing, the index of
the Dirac operator is $+3$ and hence there are 3 zero modes. There are no
other zeros of $H_w(-M)$ and hence $H_T(-M)$ until $M \approx 4$. The
largest eigenvalue of $H_T(-M)$ has a behavior as follows: Starting at
$M=0$, there is a near pairing of positive and negative eigenvalues of
the largest eigenvalue $\lambda_{\rm max}$; however, a negative
eigenvalue is slightly largest in magnitude. As $M$ increases, so does
$|\lambda_{\rm max}|$. At $M=1.249$ $\lambda_{\rm max}$ reaches $-1$,
giving a zero eigenvalue for $T^{-1}$. This corresponds to ${\tilde B}$
in the transfer matrix $T$, eqs.~(\ref{eq:Tinv_DW}) and (\ref{eq:T_DW}),
having a zero eigenvalue.
In the free field case the domain wall mass at which $T^{-1}$
has a zero eigenvalue (namely $M=1$) is identified as the mass
where the domain walls (each end of the fifth dimension) have the least
coupling for fixed $L_s$, {\it i.e.} this is the optimal domain wall mass.
As $M$ increases to the pole position of $M=2.55$, the
eigenvalue $-\lambda_{\rm max}$ diverges. The pole is the mechanism by
which the topology changes for as $M$ increases beyond the pole mass
the largest eigenvalue flips sign and decreases from positive
infinity. There is a net increase in the number of positive states of
$H_T(-M)$ occurring in $\epsilon(H_T(-M))$ for the 4-d domain wall
Dirac operator $D_{ov}(0;H_T(-M))$ and the index becomes $+4$ with
four zero modes.

In general then for the standard domain wall action in the smooth
field case with $a_5=1$, for every crossing of $H_w(-M_0)$ for $M_0<2$,
there are four opposite crossings around $M\ge 2$ and a pole of
$H_T(-M)$ at $M=2+M_0$. If at some mass $M<2$ there is an index $Q(M)$
then $Q(2+M) = -4 Q(M)$. For the overlap Dirac operator
$D_{ov}(0;H_w(-M))$, however, $Q(2+M)=-3 Q(M)$. 

Varying the fifth dimension lattice spacing $a_5$ gives some freedom
in improving the chiral symmetry properties for the induced 4-d
Dirac operators at finite $L_s$ given in
eq.~(\ref{eq:d_tov_again}).  If at some positive $\lambda_* < 1$ we have
$1-\varepsilon_{L_s/2}(a_5\lambda_*) < \delta$ where $\delta$ is some
prescribed accuracy, then the range of accuracy is
$|\lambda|$ in $[\lambda_*,1/(a_5^2 \lambda_*)]$.
For the overlap Dirac operator with
$H_w(-M)$, the flexibility of rescaling by a gauge field dependent
$a_5$ coupled with projection helped insure that all unprojected eigenvalues of
$H_w(-M)$ where within some prescribed range of accuracy of
$|\varepsilon_{L_s/2}(a_5 \lambda_i)|$ approximating $1$.

For the standard domain wall operator, $H_T(-M)$ depends implicitly on
$a_5$, and choosing different values for $a_5$ leads to different
cut-off effects, just as choosing different values for $M$ does.
Therefore, both $M$ and $a_5$ need to be kept fixed in a given simulation.
We saw in Fig.~\ref{fig:compare_dwftrf} that even for moderate
$L_s$ the largest eigenvalue $\lambda_{\rm max}$ of $H_T(-M)$ was
usually well within the good approximation region of $\epsilon(a_5
\lambda)$.
Thus we would like to use an $a_5>1$ to fully use the region of good
approximation to $\epsilon$. However, this can mix in doubler contributions,
since for each crossing at $M_0<2$ there is the doubler pole of $H_T(-M)$
at $M=2/a_5+M_0$, and as $a_5$ increases the pole moves down to lower
domain wall mass $M$. However, as can be seen in the right panel of
Fig.~\ref{fig:compare_dwftrf} the deviation for the largest
eigenvalue of $H_T$ becomes sizeable well before the pole at $M=2/a_5+M_0$
is reached. This region of large deviation moves down with increasing
$a_5$ and begins to conflict with our goal of increasing $a_5$ to
fully utilize the region of good approximation to $\epsilon$.
Since one needs to choose $M$ sufficiently larger than the $M_c$ around
which the physically important crossings, due to large instantons,
occur, the choice of $a_5$ is quite limited.
This is a fundamental limitation of the standard domain wall operator.

\subsection{Quenched gauge field}

\begin{figure}
\centerline{{\setlength{\epsfxsize}{6in}\epsfbox[0 0 576 576]{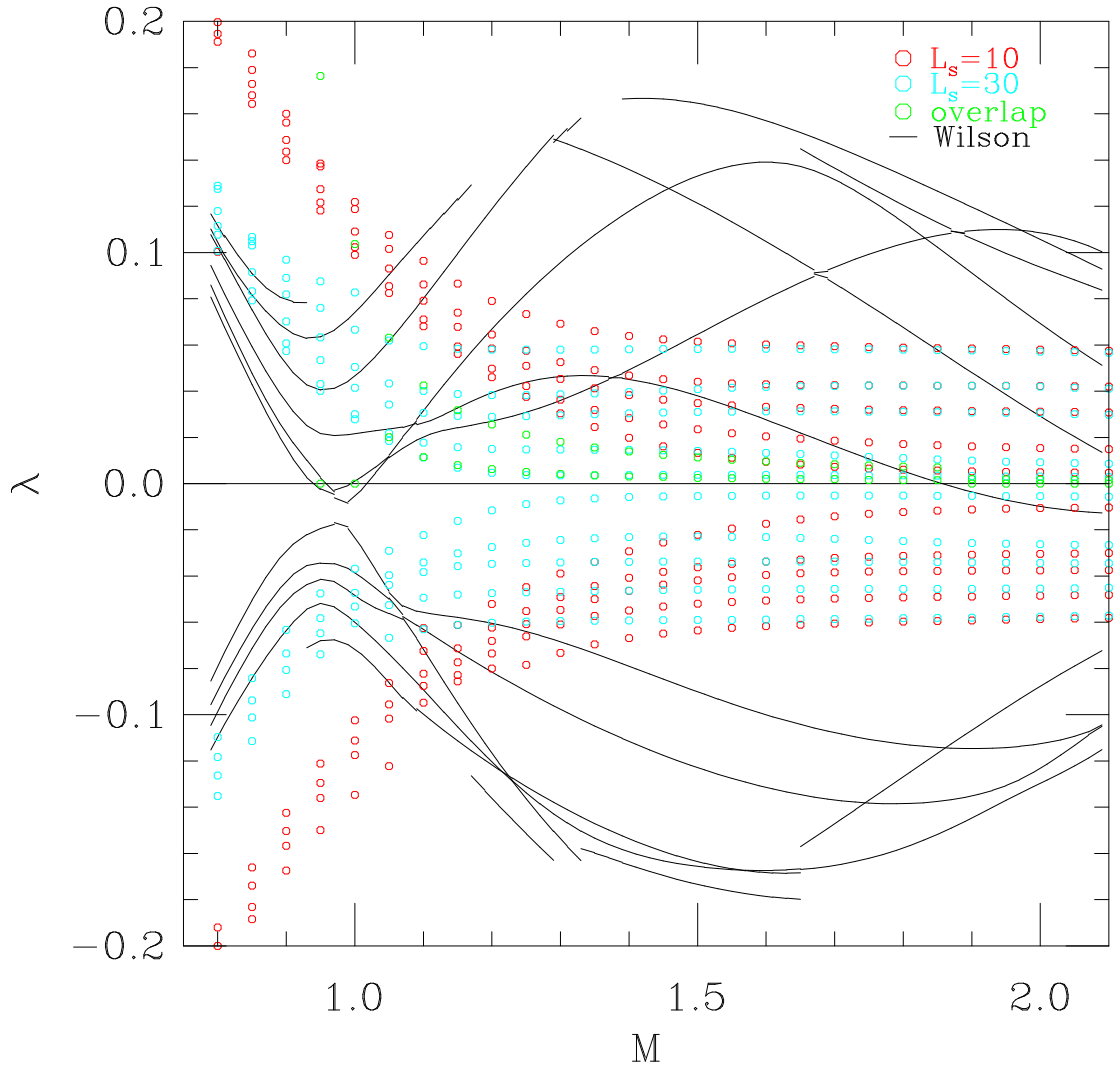}}}
\caption{Spectral flow of the hermitian (unprojected) DWF operator for
various $L_s$, and of the hermitian Wilson and overlap-Dirac operators
on a quenched SU(3), $\beta=5.85$, $8^3\times 32$ configuration around
the zero crossing of the Wilson flow.
The overlap operator correctly finds zero eigenvalues
around $M=1$ while the domain wall operator misses them.}
\label{fig:dwf_flow_cfg82}
\end{figure}

\begin{figure}
\centerline{{\setlength{\epsfxsize}{6in}\epsfbox[0 0 576 576]{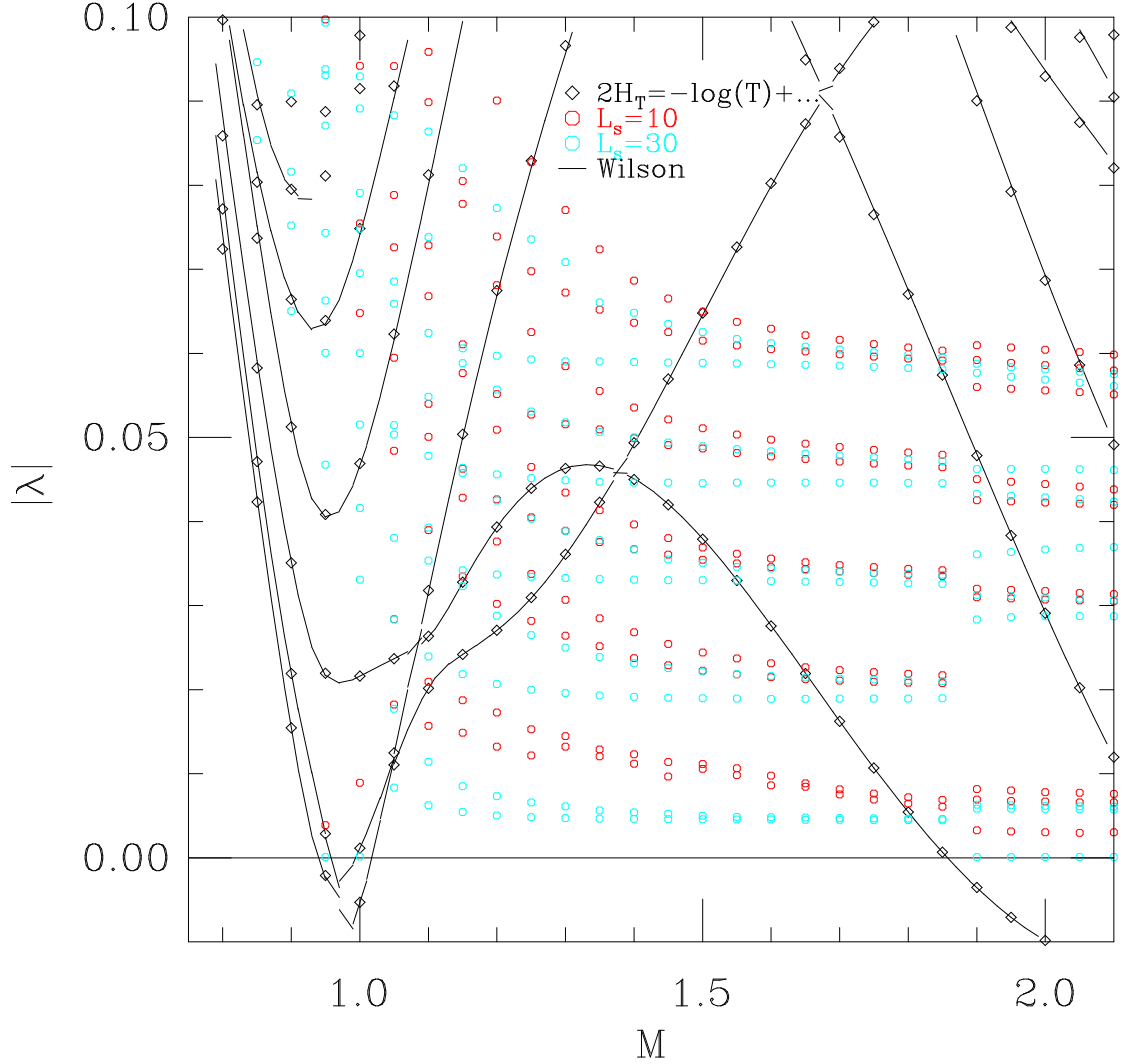}}}
\caption{Spectral flow of the projected DWF operator, projecting the
20 lowest eigenvectors of $H_T$, for various $L_s$ on the
quenched SU(3) configuration of Fig.~\ref{fig:dwf_flow_cfg82}. With
projection, the lowest domain wall eigenvalue drops to zero (within
precision used here) discontinuously after the crossing around $1.85$.
The zero eigenvalues around $M=1$ are also correctly identified.  Also
shown are the flows of $H_w$ and $2*H_T$. The eigenvalues of $2*H_T$
track closely those of $H_w$.}
\label{fig:dwf_proj_flow_cfg82}
\end{figure}

We now move on to the interesting case of a thermalized quenched SU(3)
gauge field configuration. Fig.~\ref{fig:dwf_flow_cfg82} shows the
spectral flow of the hermitian domain wall and Wilson operators for a
quenched SU(3) $\beta=5.85$, $8^3\times 32$ configuration.
Near $M=1.0$, the Wilson flow has two crossings down and two
crossings up that result in a zero index for $M>1.017$. There is a
fifth crossing at about $M_0=1.86$ resulting in an index of $-1$ for $M$
greater than this crossing mass. In the region around $M=1.0$ where
there is a non-zero index, the domain wall eigenvalues do not go
close to zero. Well into the region $M>1.017$ where the index is zero
there are fairly small domain wall eigenvalues, but no dramatic drop
of the lowest eigenvalue is observed for $M > M_0$. The overlap Dirac
operator, on the other hand, has zero modes whenever the index is non-zero.
Fig.~\ref{fig:dwf_proj_flow_cfg82} shows the spectral flow of the
projected domain wall operator on the same configuration
as in the Fig.~\ref{fig:dwf_flow_cfg82}, together with the Wilson flow
and the low-lying eigenvalues of $2H_T$. Near $M=1.0$ and
after the Wilson crossing around $1.85$ there are small (zero) eigenvalues of
the projected domain wall operator.

In the left panel of Fig.~\ref{fig:flow_cfg82} is
shown the spectral flow of the lowest eigenvalue of the standard
domain wall operator for $L_s=10$ and $30$ and the Bori\c{c}i variant
for $L_s=10$ all for $a_5=1$. We see in Fig.~\ref{fig:flow_cfg82}
that a small (near) zero eigenvalue is obtained for $L_s=30$ with the
standard domain wall $D^{(5)}_{DW}(0)$. For $L_s=10$, a drop is seen
in the lowest eigenvalue for $M>M_0$, down to the level of the
$L_s=30$ eigenvalue. For the Bori\c{c}i variant $D^{(5)}_{DW'}(0)$, a
relatively small eigenvalue is seen even for $L_s=10$. 

The explanation for the small eigenvalues of the Bori\c{c}i variant
even at $L_s=10$ can be seen in the right panel of
Fig.~\ref{fig:flow_cfg82} which shows the accuracy of the
$|\varepsilon_{L_s/2}(\lambda_i)|$ approximating $1$. The lowest
eigenvalues of $H_T(-M)$ are about half those of $H_w(-M)$. Hence for
the domain wall fermion action with $L_s=10$ and $M=1.65$ we find
$1-|\varepsilon_5(\lambda_{20})|\approx 0.25$ while for the Bori\c{c}i
variant using $H_w(-M)$ the deviation is about $0.3\%$. The largest
eigenvalue of $H_T(-1.65)$ is about $1.13$ hence the deviation is
below the range in the graph even for $L_s=10$. For the Bori\c{c}i variant, the
largest eigenvalue of $H_w(-1.65)$ is $5.82$ and there is about $6\%$
deviation. While the condition number is smaller for $H_T(-1.65)$ than
$H_w(-1.65)$, the eigenvalues are placed more optimally around one for
$H_w$ which results in overall smaller deviations of
$|\varepsilon_{L_s/2}(\lambda_i)|$ from $1$. For this configuration,
there is not much need for rescaling the Wilson $H_w$ eigenvalues by
$a_5$ since $a_5=1$ is close to optimal. For the standard domain wall
action, a larger $a_5$ would lower the deviations of
$|\varepsilon_{L_s/2}(a_5 \lambda_i)|$ from 1 for this configuration.
However, as explained in the previous sub-section, choosing an $a_5>1$
might not be desirable, since it could lead to large deviations (due
to the proximity of poles) for other configurations in a simulation.

Fig.~\ref{fig:flow_cfg82} shows that small eigenvalues of the domain
wall fermion actions $D^{(5)}_{DW}$ and $D^{(5)}_{DW'}$ can be easily
achieved using projection and the cost of the projection is
justified. To achieve the same small eigenvalue without projection
would require $L_s\gg 30$.

In Fig.~\ref{fig:gmor_cfg82} is shown the ratio $R$ defined from the
generalized Gellmann-Oakes-Renner relation (GMOR) in
(\ref{eq:GMOR_ratio}) using $M=1.65$, $a_5=1$ and the same
configuration as in Fig.~\ref{fig:flow_cfg82}. For the standard domain
wall action, not much is gained from projection for $L_s=10$ and
violations of the relation are large. For $L_s=30$, projection
significantly improves chiral symmetry resulting in small violations
of the relation while there are large deviations for even this large
$L_s$ without projection. The Bori\c{c}i variant also shows similar
improvement with increasing $L_s$. 

The condition number of a fermionic operator is a useful measure of
its convergence properties in solving linear systems of equations, and
can be used to construct a bound on the number of iterations for
convergence in some iterative methods. The condition number defined
here is $\lambda_{\rm max} / \lambda_{\rm min}$ where $\lambda_i\ge 0$
is defined from $D^{\dagger}D v_i = \lambda_i^2 v_i$.  The condition
numbers of the various fermion operators are shown in
Fig.~\ref{fig:cond}. The parameters are chosen the same as for the GMOR
test in Fig.~\ref{fig:gmor_cfg82}. The condition number of the overlap
Dirac operator $D_{ov}(m;H_w)$ is much smaller than the 5-d
methods. The condition number of the standard domain wall operator
$D^{(5)}_{DW}$ is larger for $L_s=30$ compared to $L_s=10$ while the
Bori\c{c}i variant is roughly the same for both $L_s$ values, but
higher than that of $D^{(5)}_{DW}$. Projection also does not affect
the condition numbers much basically because for this configuration
at mass $M=1.65$,
there should be no zero modes and the smallest eigenvalue of $D^{(5)}$
is not strongly affected by projection as can be seen in
Fig.~\ref{fig:flow_cfg82}. The reason the 4-d operator has much
smaller condition number is due to the largest eigenvalue which for
$D_{ov}(m)$ is very close to $1$, while for $D^{(5)}$ it is roughly
$10$.

Performance tests and comparisons of the various fermion actions are
shown in Fig.~\ref{fig:h_apps_cfg82}. The left panel shows the
iteration count for the benchmark Conjugate-Gradient for Normal
Equations (CGNE) algorithm to
achieve a residual accuracy of $10^{-5}$ normalized by the source norm
for the linear system $D^{\dagger} D \phi = D^{\dagger}b$.  The
configuration parameters are those of the GMOR test in
Fig.~\ref{fig:gmor_cfg82}. The average and the standard error of
the data (not the mean) is derived from the $12$ source color and spin
inversions necessary to compute a fermion propagator.  For the 4-d
case (exact chiral symmetry not necessary), we note that a single
inversion could be used for all the fermion masses $m$ used in
Fig.~\ref{fig:h_apps_cfg82} with the convergence governed by the time
required for the smallest fermion mass.  The 4-d overlap Dirac
operator $D_{ov}(m)$ has a low iteration count requiring about $320$
CG iterations for convergence at $m=0$. Without preconditioning, the
standard domain wall operator $D^{(5)}_{DW}(m)$ requires roughly 1000
iterations for convergence at $L_s=10$ and there is a strong $L_s$
dependence as expected from the condition number. Preconditioning
reduces the condition numbers by about a factor of three. 
The Bori\c{c}i variant
requires correspondingly more iterations for convergence. All these
results are consistent with a linear scaling in the condition number.

It has been pointed out~\cite{Borici,Borici_thesis} that both
CGNE and Conjugate-Residual are optimal algorithms for
the overlap-Dirac operator with exact chiral symmetry. In
principle the Conjugate-Residual method would be more desirable since
only one $D_{ov}(m)$ application is needed per iteration. For CGNE
one expects two applications of $D_{ov}(m)$, but this can be rewritten
as only one application of $\epsilon(H_w)$ on a vector~\cite{EHN_cond},
so the work is the same per iteration for both methods. However, it was
found that Conjugate-Residual had clearly worse convergence in practice
compared to CGNE and was not used.

The right panel of Fig.~\ref{fig:h_apps_cfg82} shows the actual cost
for convergence measured in total number of $D_w$ applications for
each linear system solution. This includes the ``inner'' CG portion
for the 4-d overlap Dirac operator. For each ``outer'' CG iteration, a
single application of $\epsilon(H_w)$ is needed on a
vector~\cite{EHN_practical}. This inner CG iteration in turn involves
multiplications of a vector with $H_w^2$. For the standard domain wall
operator, there are $2 L_s$ applications of $D_w$ needed per
iteration. For the Bori\c{c}i variant, there are $3 L_s$ applications
of $D_w$ needed as can be seen from the form of $D^{(5)}_{DW'}(m)$ and
$(D^{(5)}_{DW'}(m))^\dagger$ in eqs.~(\ref{eq:D_5d_matrix}) and
(\ref{eq:Ddag_5d_matrix}).

The cost of the methods can be roughly characterized as follows: the
Bori\c{c}i variant requires about twice as many $D_w$ applications for
a given $L_s$ compared to the standard domain wall operator with and
without projection. This can be expected from the slightly larger
condition number (mainly from the largest eigenvalue) and the 50\%
overhead in $D_w$ applications. Preconditioning saves the domain wall
operator roughly a factor of two to three for both $L_s$ values with and
without projection. This decrease is mainly due to the reduction of
the largest eigenvalue.
The $L_s=30$ calculations require about three to
four times as many $D_w$ applications as the $L_s=10$ calculations
with and without projection. However, $L_s=30$ and projection is
needed to pass the GMOR test successfully.

The inversion of $(1-{\hat A}_{ee} P_-)$ in eq.~(\ref{eq:precond})
requires only about three iterations of a Minimal Residual algorithm to
reach an accuracy of $10^{-7}$. This multiplicative overhead in the
number of ``outer'' CG iterations, but not proportional to $L_s$, is
included in the total cost of the preconditioned domain wall operator
and results in about an 18\% overhead for $L_s=10$ and a 5\% overhead for
$L_s=30$ --- there is little $L_s$ dependence in the ``inner'' CG.
However, for these parameters and configuration (topological index is
zero) the projected preconditioned domain wall requires fewer CGNE
iterations for convergence, and is less costly in $D_w$ applications
for $L_s=30$ than the unprojected version.

For $L_s=30$ with preconditioning and projection, the standard domain
wall operator is about three times less costly in $D_w$ applications
than the 4-d overlap Dirac operator.  However, given that with the
4-d overlap Dirac operator all the
masses for this spectroscopic calculation can be computed
simultaneously for a given color and spin source, the cost benefit of
the two approaches is comparable.

\section{Discussion}
\label{sec:discussion}

The five dimensional domain wall fermion action provides a means
whereby an effective chiral theory may be obtained. Upon integrating
out extra modes (which may be interpreted as flavors) from this extra
fifth dimension, a chiral theory is obtained in four dimensions. For
finite fifth dimensional extent $L_s$, the form of this action in four
dimensions is
\begin{eqnarray}
D_{tov}(m; H_T) = 
\frac{1}{2} \Bigl[1+m + (1-m) \gamma_5 \varepsilon_{L_s/2}(a_5 H_T) \Bigr]
\label{eq:d_tov_again2}
\end{eqnarray}
with the auxiliary Hamiltonian $H_T$. When $L_s\rightarrow\infty$,
$\varepsilon_{L_s/2}(a_5 H_T)\rightarrow \epsilon(H_T)$ for $a_5>0$
and there is a chiral symmetry in the action.  For finite $L_s$,
chiral symmetry breaking is induced because
$\varepsilon_{L_s/2}(a_5\,H_T)$ deviates from one.  These deviations
occur whenever the range of eigenvalues of $H_T$ are outside the range
of approximation of $\varepsilon_{L_s/2}(x)$ to
$\epsilon(x)$. Deviations can occur for both large and small
eigenvalues of $H_T$.

Section~\ref{sec:PDWF} shows how the domain wall operator and a
variant can be straightforwardly modified to allow arbitrarily precise
chiral symmetry at finite values of $L_s$. The domain wall operator as
shown in eq.~(\ref{eq:D_5d_matrix}) has only two new entries
corresponding to projection of eigenvalues of $H_T$ or $H_w$.  When the
$\chi$ basis is chosen for the fermions fields in the action,
eq.~(\ref{eq:S_5d3}), the projection term occurs only in the
origin along the fifth dimensional line (the flavor index). With this
special position, the projection term is unmodified by the successive
integrations of the extra flavor fermion fields. After extraction of
the pseudo-fermion term, the expected form of the 4-d
fermion action is obtained with the projection term acting as (part)
of the spectral decomposition of $\epsilon(a_5 H)$. In principle then
one can see that once in the $\chi$ basis the projection term could
account for the entire spectrum of $H$ and the fifth dimensional
extent could be set to one!

All the normal relations involving the fermion propagator go through
with projection, and we have the relation connecting the 5-d and 4-d
operators in eq.~(\ref{eq:op5d4d}) and the propagator in
eq.~(\ref{eq:4d5d_prop}) valid with and without projection.

The main purpose of this paper is to show how projection can be used
to achieve exact chiral symmetry at finite fifth dimensional
extent. The method is not limited to quenched calculations and methods
were outlined in Section~\ref{sec:dynamical} for use in dynamical
fermion calculations. Even-odd preconditioning is practical and for
moderate $L_s$ is a small overhead.
Tests of chiral symmetry were made
for a smooth gauged field background composed of a single instanton,
and for a quenched SU(3) gauge field background. Small (zero)
eigenvalues of the Dirac operator can be achieved given careful
control of the range of the approximation of the eigenvalues of the
auxiliary Hamiltonian and choice of $L_s$. A stringent simple test is
the generalized Gellmann-Oakes-Renner relation (GMOR) from
eq.~(\ref{eq:GMOR}) and projection appears essential to pass this
test.

However, for practical simulations the requirement of no induced
chiral symmetry breaking may not be necessary or easy to achieve.  In
this case, the amount of projection and the fifth dimensional length
can be tuned. This may well be necessary anyway since as shown for
quenched backgrounds, there is evidence~\cite{EHN_rho0} for a non-zero
density of zero eigenvalues $\rho(0;H)$ of the auxiliary Hamiltonian
$H_w$ (and hence $H_T$) for a large range of $\beta$. If one wants to
go to very large lattice volume, the number of eigenvalues for
projection to achieve some given accuracy for $\varepsilon_{L_s/2}(a_5 H)$
grows like the 4-d lattice volume. For example, this cost is an
additional overhead for a spectroscopy calculation which also grows
like the 4-d lattice volume.  One way to lessen the overhead is to use
a weaker coupling since $\rho(0;H)$ decreases very rapidly with the
coupling which of course may require a larger lattice volume to hold
the physical volume fixed. To keep calculation costs down, $L_s$ can
be chosen suitably small and measures of induced chiral symmetry
breaking like PCAC~\cite{dwf_problems} can be used. 
No effort was made to quantify this for given parameters like $\beta$,
however, some projection is expected to reduce the induced chiral
symmetry breaking from finite $L_s$.

For a given calculation, one would like to know what ultimately is the
most efficient method to use. Namely, is the 5-d or the direct 4-d
method the most efficient? From the present work, when one enforces
chiral symmetry there is not really a huge difference in the cost of
any of the methods tested. Quite likely then the answer depends on
what one wants to do and on ones taste. For eigenvalue calculations of
spectral quantities, the 4-d eigenvalues are needed. For spectroscopy
or dynamical fermion calculations, either 5-d or 4-d methods can be
used.  For fixed cutoff dependence, the 5-d domain wall operator is
preferable to the 4-d methods using $\epsilon(a_5 H_T)$ -- the
nontrivial cost of applying $H_T$ within an inner CG iteration inside of
another inner CG iteration is prohibitive. For 4-d eigenvalue
calculations the application of $D_{ov}(m;H_T)$ from $D^{(5)}_{DW}(m)$
in eq.~(\ref{eq:op5d4d}) is probably most efficient.  For the overlap
Dirac case using $D_{ov}(m;H_w)$, the Bori\c{c}i variant of
$D^{(5)}_{DW'}(m)$ does not appear competitive for spectroscopy
calculations. A multi-grid implementation in the fifth direction
might help~\cite{Borici}, but has not been tested here.
Other 5-d methods should be tried and tested~\cite{HN_bounds}.

For a direct comparison of the standard domain wall action
$D^{(5)}_{DW}(m)$ and the overlap operator $D_{ov}(m)$, one first must
answer just how important is chiral symmetry? Projection is essential
for realizing good chiral symmetry properties for small quark masses
and moderate $L_s$. Clearly computing eigenvectors using $H_w$ is much
more efficient than computing eigenvectors of $H_T$. If one enforces
chiral symmetry the un-preconditioned domain wall operators is about
the same cost in the number of $D_w$ applications as the overlap
operator. Preconditioning saves about a factor of three in cost,
however the methods are comparable in cost for a spectroscopy
calculation when multiple fermion masses are needed.
A benefit of the 5-d method is that it gives a simple handle $L_s$ in
which to reduce the overall cost while incurring some acceptably small
amount of induced chiral symmetry breaking.

RGE was supported by DOE contract DE-AC05-84ER40150 under which the
Southeastern Universities Research Association (SURA) operates the
Thomas Jefferson National Accelerator Facility (TJNAF).  UMH was
supported in part by DOE contract DE-FG05-96ER40979.  Computations
were performed on FSU's QCDSP, operated at TJNAF, and on the HPC
workstation cluster at TJNAF and the workstation cluster at CSIT.  We
thank Rajamani Narayanan for discussion, and UMH would like to thank
Artan Bori\c{c}i for a useful conversation.

\newpage

\begin{figure}
\centerline{{\setlength{\epsfxsize}{6in}\epsfbox[0 0 576 576]{dwf_flow_inst.ps}}}
\caption{Spectral flow of the hermitian DWF operator for various $L_s$ and
of the hermitian Wilson operator around its zero crossing
on a single $8^4$ instanton background. Also shown is the flow of the
overlap-Dirac operator which has zero eigenvalues
after the crossing, setting in discontinuously.}
\label{fig:dwf_flow}
\end{figure}

\begin{figure}
\centerline{{\setlength{\epsfxsize}{6in}\epsfbox[0 0 576 576]{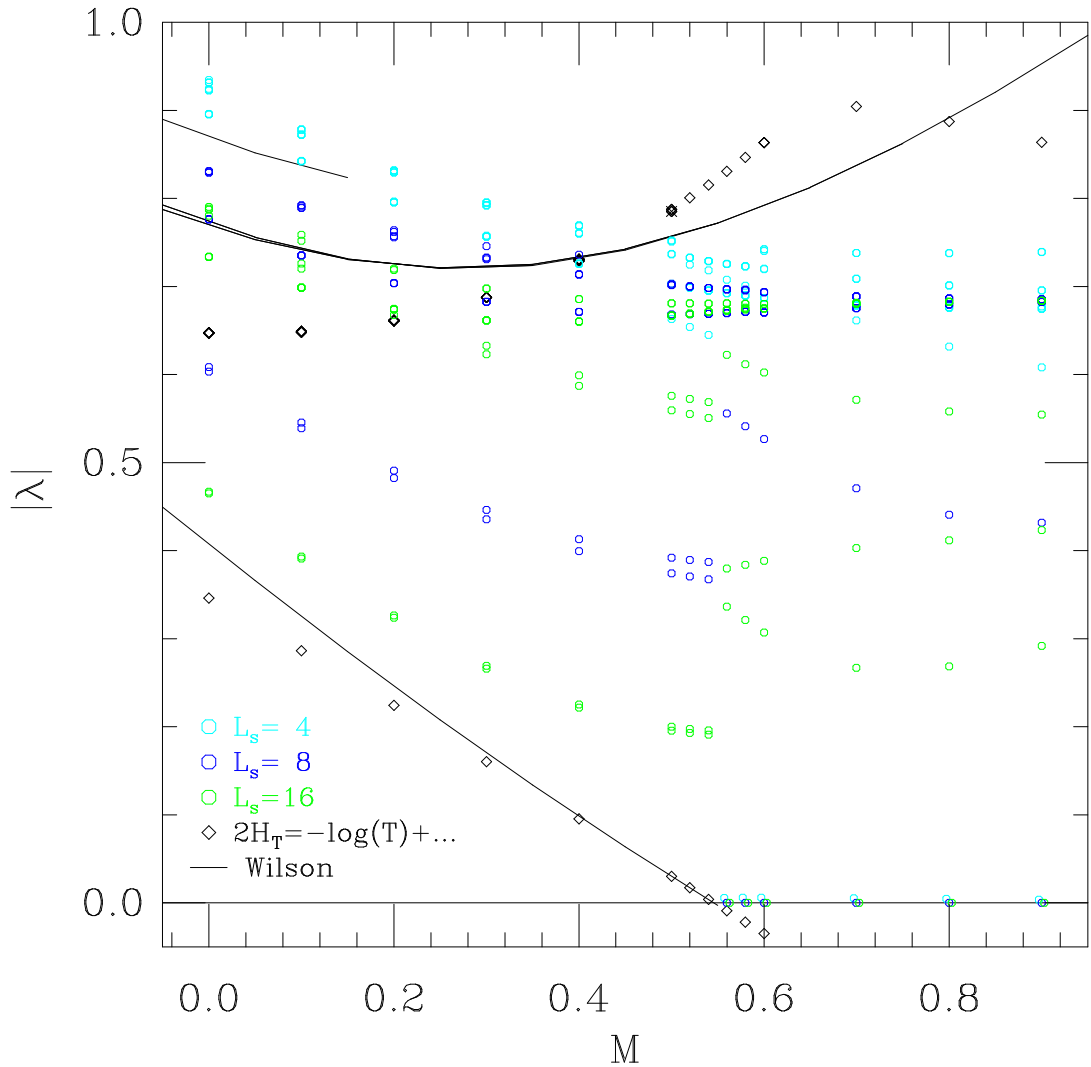}}}
\caption{Spectral flow of the projected DWF operator, projecting the
five lowest eigenvectors of $H_T$, for various $L_s$ on the same
instanton background as in Fig.~\ref{fig:dwf_flow}. With projection,
the lowest DWF eigenvalue drops to (near) zero discontinuously after
the Wilson crossing.  Also shown are the flows of $H_w$ and $2*H_T$.
The eigenvalues of $2*H_T$ track closely those of $H_w$.}
\label{fig:dwf_proj_flow}
\end{figure}

\begin{figure}
\centerline{{\setlength{\epsfxsize}{6in}\epsfbox[60 150 576 400]{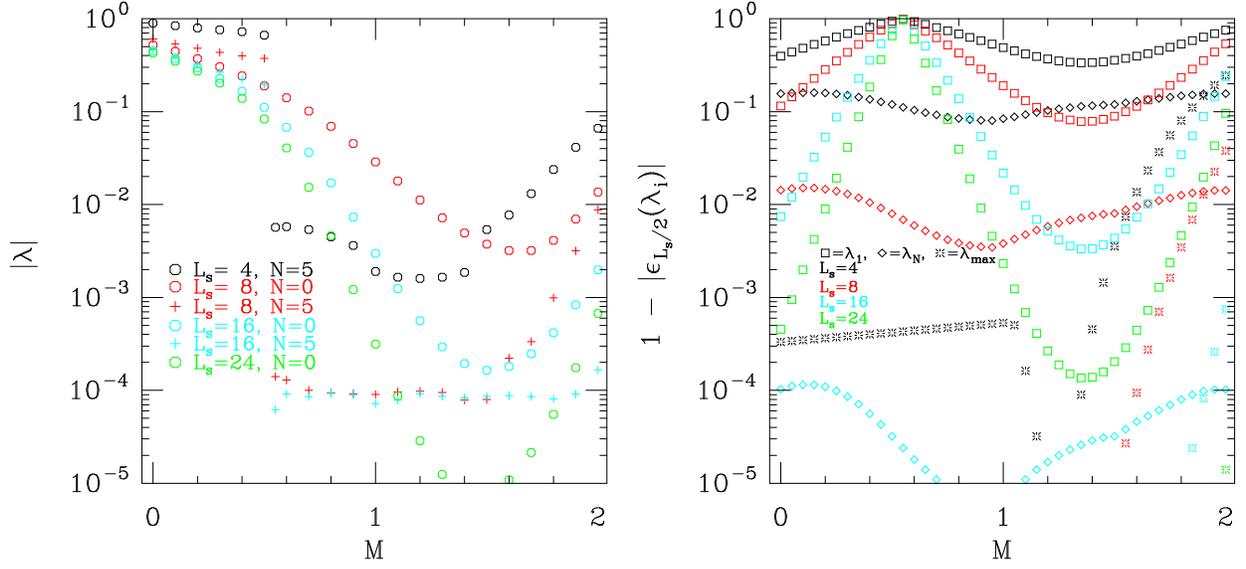}}}
\caption{Results for DWF on the same $8^4$ instanton background as in
in Fig.~\ref{fig:dwf_flow}. The left panel shows a logarithmic plot of
how the lowest (near) zero eigenvalue (with extent $L_s$ and $N$
eigenvectors projected) slowly rises with decreasing accuracy of the
epsilon approximation as shown on the right.  The right panel shows
the quality of the approximation of $|\epsilon_{L_s/2}(\lambda_i)|$ to
unity for eigenvalues $\lambda_i$ of $H_T$. The $\lambda_1$ deviations
shows the quality for the standard action, $\lambda_N$ is for the
fifth eigenvalue, and $\lambda_{max}$ is for the highest
eigenvalue. The smallest projected domain wall eigenvalue rises as the
deviation for $\lambda_{max}$ increases.}
\label{fig:compare_dwftrf}
\end{figure}

\begin{figure}
\centerline{{\setlength{\epsfxsize}{6in}\epsfbox[0 0 576 576]{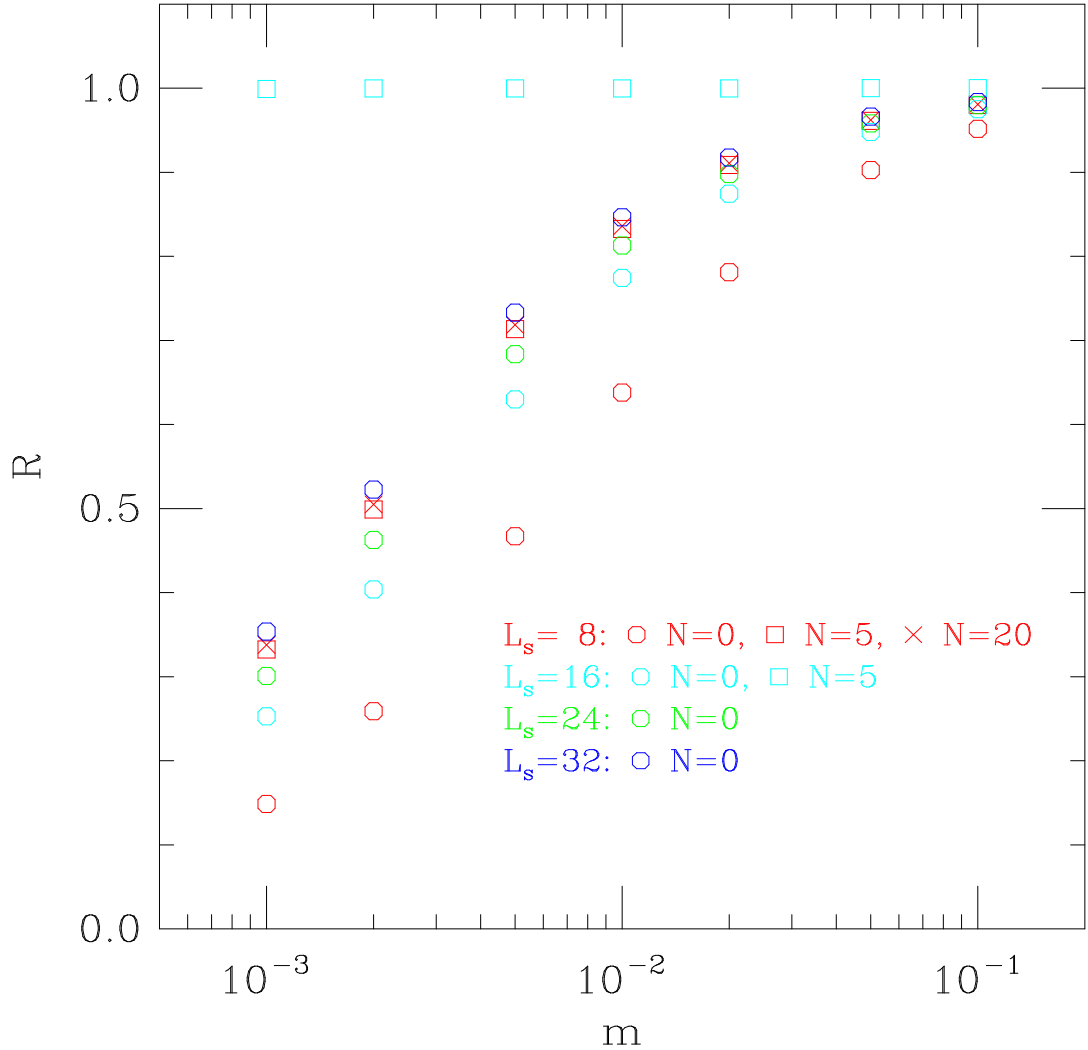}}}
\caption{The ratio given by the Gellmann-Oakes-Renner relation (GMOR)
for $M=0.5$ (just before the crossing) as a function of $L_s$ and
projection on a single $8^4$ instanton background. $N$ denotes the
number of projected eigenvectors. Projection for $L_s=16$ helps to
satisfy GMOR since for all unprojected eigenvalues of $H_T$,
$\epsilon(H_T)$ is well approximated. For $L_s=8$, the deviation for
the largest projected eigenvalue is still large as seen in
Fig.~\ref{fig:compare_dwftrf}.}
\label{fig:smooth_GMOR}
\end{figure}

\begin{figure}
\centerline{{\setlength{\epsfxsize}{6in}\epsfbox[60 150 576 400]{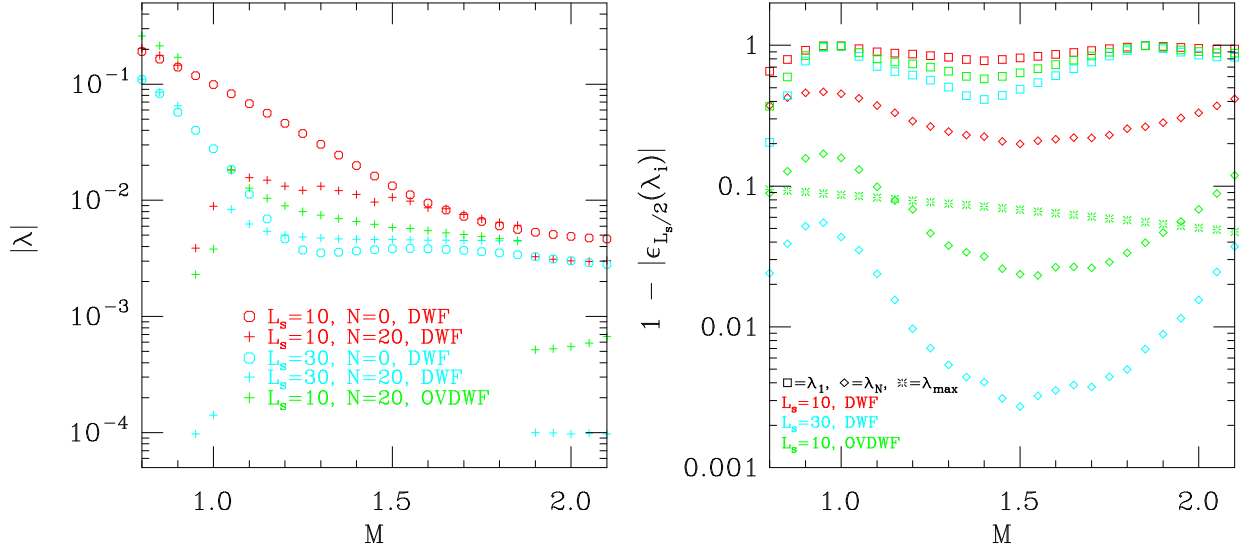}}}
\caption{Results for the quenched SU(3) configuration of
Fig.~\ref{fig:dwf_flow_cfg82}. The presentation is similar to
Fig.~\ref{fig:compare_dwftrf}. DWF and OVDWF are the domain wall operator and
Bori\c{c}i's variant, resp., with $N$ eigenvectors
projected. The left panel shows how the lowest (near) zero eigenvalue
slowly rises with decreasing accuracy of the epsilon approximation as
shown on the right.  The right panel shows the quality of the
approximation of $|\epsilon_{L_s/2}(\lambda_i)|$ for eigenvalues
$\lambda_i$ of $H_T$. Here $N=20$ eigenvectors are projected. For
domain wall, the 20-th eigenvalue is badly approximated for $L_s=8$
but slightly better for the Bori\c{c}i variant. For $L_s=30$, the
deviation of $\lambda_{\max}$ is below the range shown.}
\label{fig:flow_cfg82}
\end{figure}

\begin{figure}
\centerline{{\setlength{\epsfxsize}{6in}\epsfbox[0 0 576 576]{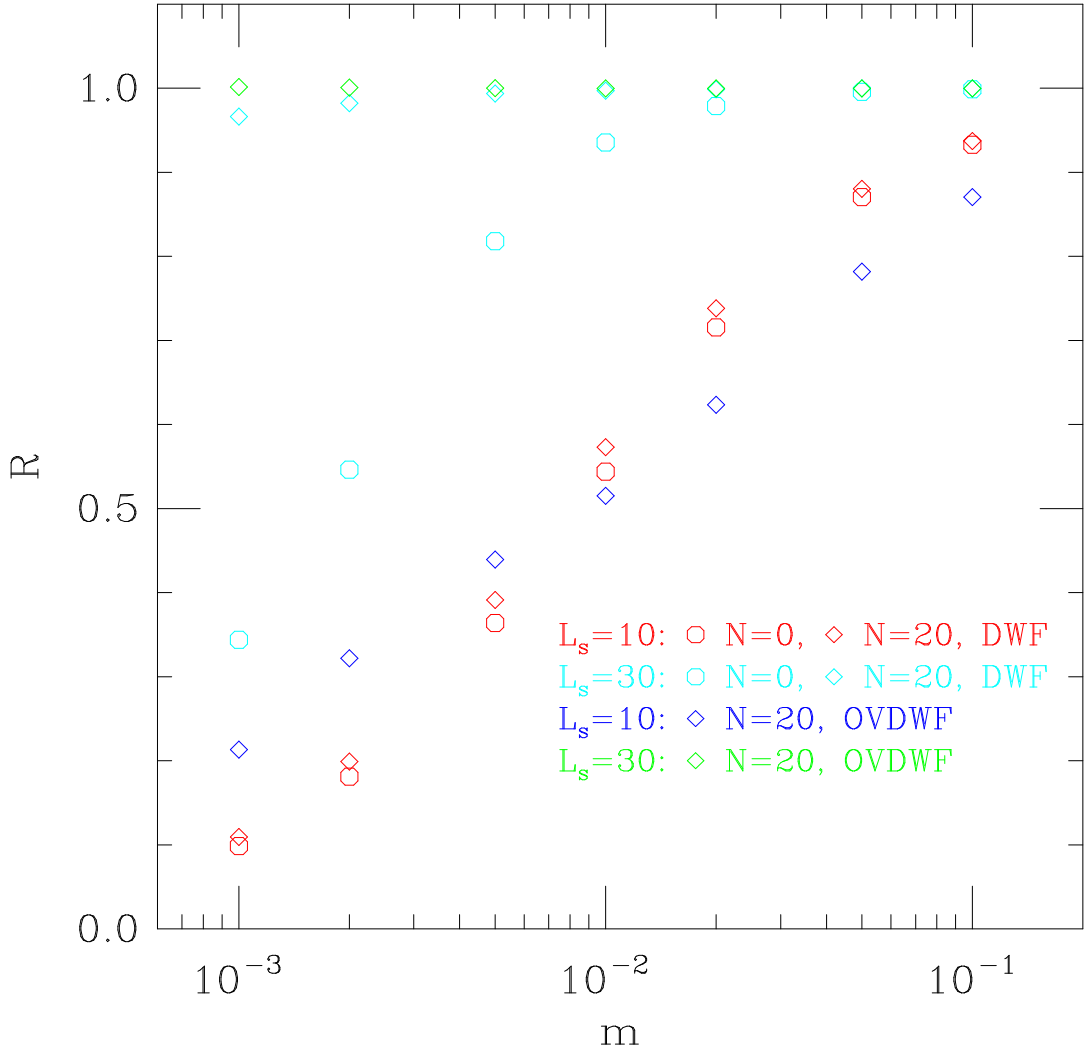}}}
\caption{The ratio given by the Gellmann-Oakes-Renner relation (GMOR,
c.f.~(\ref{eq:GMOR})) for $M=1.65$ (just before the crossing) as a
function of $L_s$ and projection on the same quenched SU(3)
configuration as in Fig.~\ref{fig:dwf_flow_cfg82}. DWF and OVDWF are
the domain wall operator and Bori\c{c}i's variant, resp. For $L_s=30$ with
projection, the GMOR relation is well satisfied but some discrepancy
is seen from deviations of $H_T$ eigenvalues $\lambda_i$ in
Fig.~\ref{fig:flow_cfg82}.}
\label{fig:gmor_cfg82}
\end{figure}

\begin{figure}
\centerline{{\setlength{\epsfxsize}{6in}\epsfbox[0 0 576 576]{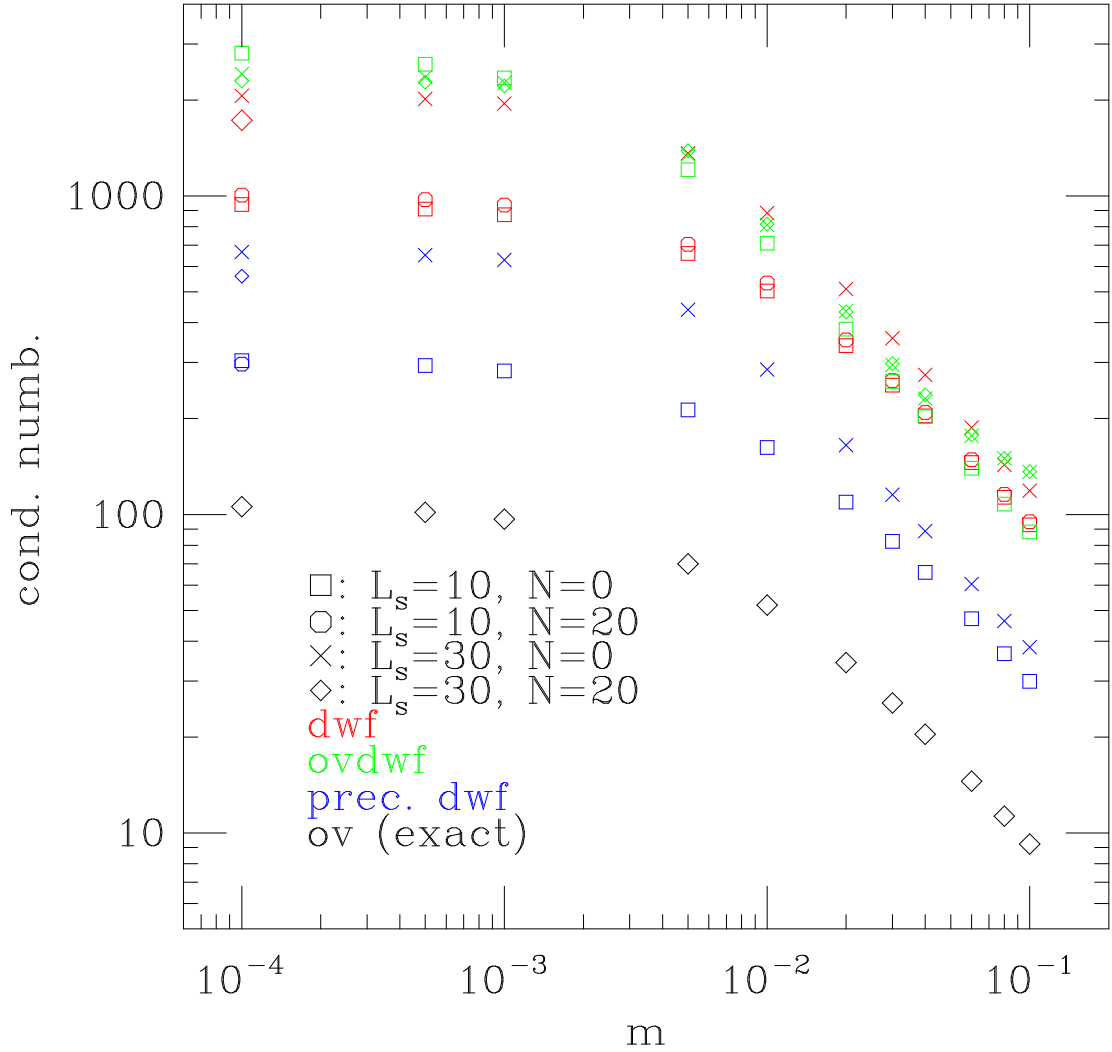}}}
\caption{Condition numbers as a function of the mass for various actions
on the same quenched SU(3) configuration and parameters as in
Fig.~\ref{fig:gmor_cfg82}. The overlap operator has the lowest
condition number.}
\label{fig:cond}
\end{figure}

\begin{figure}
\centerline{{\setlength{\epsfxsize}{6in}\epsfbox[40 150 576 400]{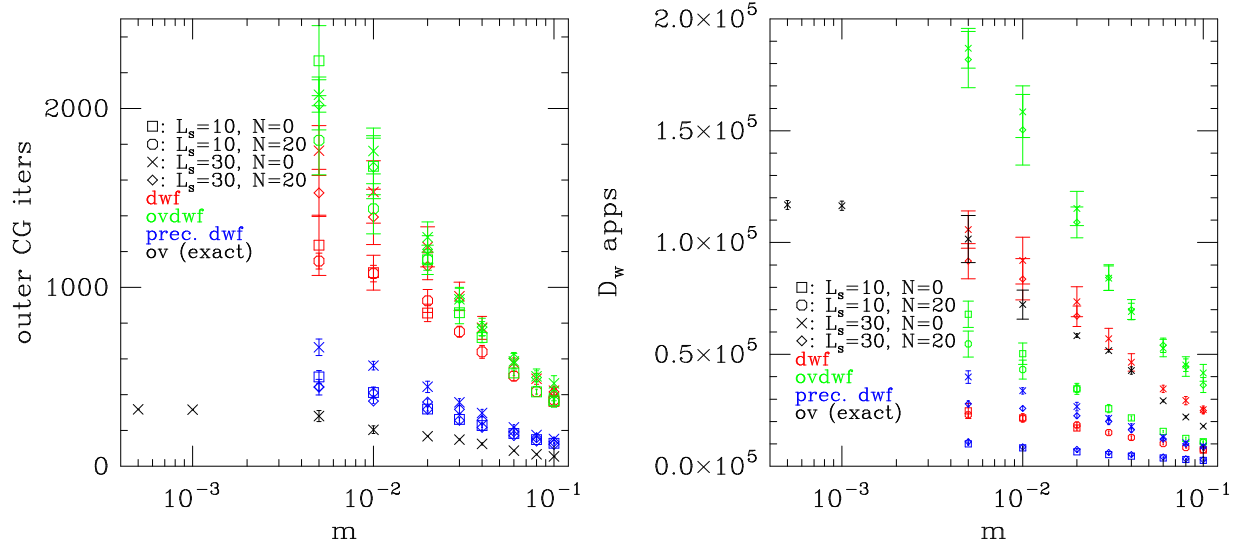}}}
\caption{The left panel shows the number of (outer) CGNE applications
to reach convergence of $10^{-5}$.  The right panel shows the total
number of $D_w$ applications for convergence (including inner and outer
CG applications) on the same quenched SU(3) configuration and
parameters as in Fig.~\ref{fig:gmor_cfg82}. The overlap operator needs
the fewest CG iterations. For $L_s=30$ with projection, the standard
(un-preconditioned) domain wall and overlap operator have comparable
$D_w$ application counts. Preconditioning saves the domain wall
operator about a factor of two to three in cost.}
\label{fig:h_apps_cfg82}
\end{figure}

\end{document}